\documentclass[preprint,11pt]{elsarticle}

\usepackage{hyperref}

\usepackage{multirow}

\usepackage[ruled,lined,algonl,boxed]{algorithm2e}

\usepackage[table]{xcolor}

\usepackage{amsmath,amssymb}

\usepackage{amsthm}

\allowdisplaybreaks

\newdefinition{rmk}{Remark}
\newproof{pf}{Proof}
\newproof{pot}{Proof of Theorem \ref{thm2}}

\journal{Journal}

\usepackage{amsopn}
\usepackage{float}
\usepackage{xcolor, graphicx}
\usepackage{multirow}
\usepackage{hhline}
\usepackage{makecell}
\usepackage{amsmath, amssymb, amsthm}
\usepackage[a4paper,total={137mm,210mm}]{geometry}

\newtheorem{theorem}{Theorem}
\renewcommand\theadfont{}
\DeclareMathOperator*{\argmin}{arg\,min}
\DeclareMathOperator*{\argmax}{arg\,max}
\begin{document}

\begin{frontmatter}



\title{Greedy and randomized heuristics for optimization of\\ $k$-domination models in digraphs and road networks}

\author[label2]{Lukas Dijkstra}%\corref{cor1}}
\ead{dijkstral@cardiff.ac.uk}
%%%%
\author[label2]{Andrei Gagarin}%\corref{cor1}}
\ead{gagarina@cardiff.ac.uk}
%%%%%
\author[label2a]{Padraig Corcoran}
\ead{corcoranp@cardiff.ac.uk}
\author[label2]{Rhyd Lewis}%\corref{cor1}}
\ead{lewisr9@cardiff.ac.uk}
\address[label2]{School of Mathematics, Cardiff University, Cardiff, Wales, UK}
\address[label2a]{School of Computer Science \& Informatics, Cardiff University, Cardiff, Wales, UK}

%%%%%%%%%%%%%%%%%%%%%%%%%%%%%%

\begin{abstract}
Directed graphs provide more subtle and precise modelling tools for optimization in road networks than simple graphs.
In particular, they are more suitable in the context of alternative fuel vehicles and new automotive technologies, like electric vehicles. 
In this paper, we introduce the new general 
concept of a reachability digraph associated with a road network 
to model the placement of refuelling facilities in road networks as $k$-dominating sets in the reachability digraph. 
Two new greedy heuristics are designed and experimentally tested to search for small $k$-dominating sets in two types of digraphs,
including the reachability digraphs. 
Refined greedy strategies are shown to be efficient, capable of finding good quality solutions, and 
suitable for application in very large  
digraphs and road networks.
Also, a probabilistic method is used to prove a new upper bound on the $k$-domination number of a digraph, 
which informs the development of a new 
randomized heuristic to search for $k$-dominating sets in the digraph.
Generalizing the randomized heuristic ideas, making the heuristic more flexible, tuning and combining it with the greedy strategies allows us to 
obtain even better results for the reachability digraphs. 
Computational experiments are conducted for a case study of road networks in the West Midlands (UK).
\end{abstract}

\begin{keyword} 
Digraph models \sep Optimization in road networks \sep Facility location problems \sep $k$-Domination in digraphs \sep Heuristics 


\end{keyword}

\end{frontmatter}

%%%%%%%%%%%%%%%
%%%%%%%%%%%%%
\section{Introduction}
\label{sec-intro}

Graphs and the tools of graph theory are widely used to model and solve many modern optimization problems,
particularly, in road networks \cite{BH15, CG2021, FNS2015, GC2018, Zve2}.
Different types of dominating sets in graphs are often considered as general modelling tools for facility location problems \cite{CH1977, HHS1998, Zve2, WDC11}.
They have attracted a lot of attention in the literature from the theory point of view \cite{AS1992,CH1977, HHS1998}
and in applications \cite{GC2018, Zve2, WDC11, YW2013}.
However, usually dominating set problems are formulated and considered using simple graphs,
and little attention has been paid to more general problem modelling tools like digraphs.
For instance, the two classic books on dominating sets \cite{HHS1998, HHS1998-2} contain only one chapter on digraphs,
and the classic book on digraphs \cite{BG2009} pays limited attention to dominating set models.  

In the context of road networks, it can be seen that simple graph models are quite limited and, in many cases, ambiguous.
For example, they are not capable of modelling one-way streets,
as well as the fact that vehicles use more fuel and energy when going uphill. 
In fact, most electric vehicles (EV's) can generate energy when going downhill (e.g., see regenerative braking in \cite{CMC2010, DM2017}), 
as opposed to energy consumption.
This can result in a negative travelling cost in one direction and a high positive travelling cost in the opposite direction on the same road segment. 
Thus, despite the fact that important optimization problems, such as those in road networks, are best modelled on digraphs,
some facility location problems and their potential solution methods are usually not considered on this type of graph. 

In this paper, we address the problem of modelling and optimizing facility locations in road networks using directed graphs (digraphs). 
We propose and illustrate novel  
efficient and effective search heuristics to deal with the small size $k$-dominating set problem in digraphs in general,
and  experimentally test the heuristic solution methods on two different types of digraph, including those corresponding to road networks.\footnote{Preliminary 
results of this research are reported in the conference paper \cite{INOC2024}.}
Because of interest in this kind of problems in very large  
road networks (e.g., see \cite{FNS2015}),
in addition to efficiency and effectiveness, we also aim at simplicity and reproducibility of the designed solution methods. 
Therefore, we focus on development of generic greedy and randomized heuristic ideas 
that are reasonably simple, efficient, and eventually can be used in a more specific context.
In addition to the design of modelling tools and solution methods on digraphs, 
we explore frontiers of tractability of the new greedy and randomized heuristics for random digraphs and road network digraphs.

%Definitions \& Notations
\subsection{Digraphs}

A \emph{digraph} $D=(V_D,A_D)$ is a set of elements $V_D=\{v_1,v_2,...,v_n\}$, called \emph{vertices} of $D$,
and a set $A_D=\{e_1,e_2,...,e_m\}$ of \emph{ordered} pairs of distinct vertices of $D$, called \emph{arcs}.
So, an arc $e\in A_D$ is an ordered pair $e = (v_i,v_j)$ for some vertices $v_i,v_j \in V_D$, $i,j \in [n]=\{1,2,\ldots,n\}$, $i\not= j$, as opposed to unordered pairs of vertices in simple graphs.
Two vertices $u,v\in V_D$ are \emph{adjacent} in $D$ if $(u,v)\in A_D$ or $(v,u)\in A_D$.
Notice that a simple graph $G=(V_G,E_G)$, $V_G=\{v_1,v_2,...,v_n\}$, $E_G=\{e_1,e_2,...,e_{m'}\}$, 
where $E_G$ is a set of \emph{unordered} pairs of distinct vertices of $V_G$ called \emph{edges},
can be considered as a particular type of a digraph.
This is because each unordered pair $e=\{u,v\}\in E_G$, $u,v \in V_G$, 
can be considered as two arcs $(u,v)$ and $(v,u)$ in the corresponding digraph $D=(V_D,A_D)$, $V_D=V_G$, so that $|A_D|=2|E_G|$.

Given a vertex $v$ of a digraph $D=(V,A)$, 
the \emph{out-neighbourhood} of $v$ in $D$ is the set $N^+(v) = \{u \in V \mid (v,u) \in A\}$ of vertices adjacent to $v$ by using arcs leaving $v$ in $D$, i.e. directly reachable from $v$ in $D$,
whereas the \emph{in-neighbourhood} of $v$ in $D$ is the set $N^-(v) = \{u \in V \mid (u,v) \in A \}$ of vertices adjacent to $v$ by using arcs entering $v$ in $D$, i.e. from which $v$ can be directly reached in $D$. 
The \emph{closed out-} and \emph{in-neighbourhoods} of $v$ in $D$ are respectively $N^+[v] = N^+(v) \cup \{v\}$ and $N^-[v] = N^-(v) \cup \{v\}$. 
The \emph{out-degree} of $v$ is denoted by $d^+_v = d^+(v) = |N^+(v)|$, and the \emph{in-degree} of $v$ is denoted by $d^-_v = d^-(v) = |N^-(v)|$. 

The \emph{minimum out-} and \emph{in-degrees} of $D$ are respectively the smallest values of $d^+_v$ and $d^-_v$, $v\in D$, denoted by $\delta^+ = \delta^+(D)$ and $\delta^- = \delta^-(D)$.
The \emph{maximum out-} and \emph{in-degrees} of $D$ are the largest values of $d^+_v$ and $d^-_v$, $v\in D$, denoted by $\Delta^+ = \Delta^+(D)$ and $\Delta^- = \Delta^-(D)$, respectively.
Suppose the vertices of $D$ are labelled $v_1,v_2,\ldots,v_n$ in such a way that their in-degrees form a non-decreasing sequence $(d^-_{1},d^-_{2},\ldots,d^-_{n})=(d^-_{v_1},d^-_{v_2},\ldots,d^-_{v_n})$, $n=|V|$, 
where $d^-_{i}\leq d^-_{j}$ for any $i,j \in [n]$, $i<j$.
Then the \emph{median in-degree} of $D$ is defined as $\displaystyle d^-_{\mathrm{med}}=d^-_{\frac{n+1}{2}}$, if $n$ is odd,  and $\displaystyle d^-_{\mathrm{med}}=\frac{d^-_{\frac{n}{2}} + d^-_{{\frac{n}{2}}+1}}{2}$, if $n$ is even.
The \emph{average out-} and \emph{in-degrees} of $D$ are defined as\ \ $\displaystyle d^+_{\mathrm{ave}}=\frac{1}{n}\sum_{v\in V} d^+_v$\ \ and\ \ $\displaystyle d^-_{\mathrm{ave}}=\frac{1}{n}\sum_{v\in V} d^-_v$,\ \ respectively ($d^+_{\mathrm{ave}} = d^-_{\mathrm{ave}}$). 
For other basic terminology related to graphs and digraphs, the reader is referred to the monograph on digraphs \cite{BG2009}. 

%%%%%%%%%% definition of the reachability digraph %%%%%%%%
\subsection{Reachability digraphs for vehicles in road networks} 
\label{sec:reachability}

\begin{figure}%
\centering
\parbox{2.5in}{\includegraphics[width=8.1cm]{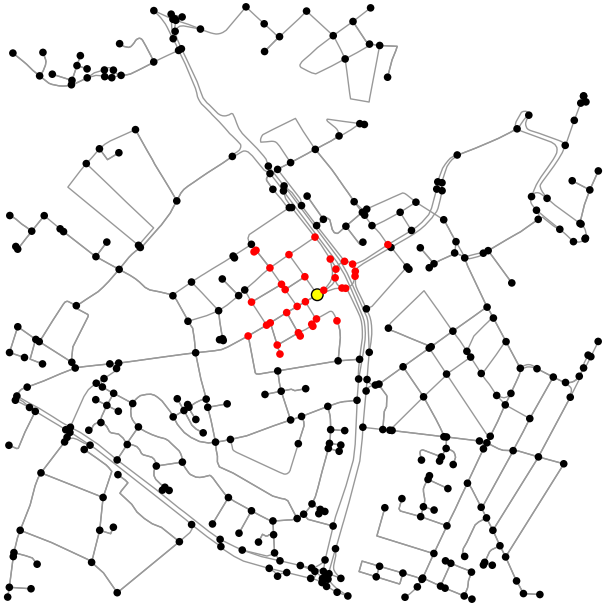}}%
\qquad 
\begin{minipage}{2.5in}%
\hspace{1.7cm} \includegraphics[width=7.3cm]{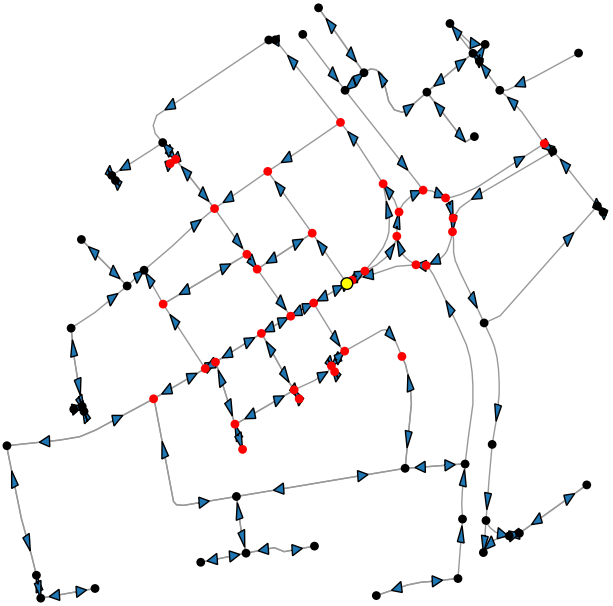}
\end{minipage}%
\caption{A digraph model of a road network showing a vertex (yellow) and its out-neighbours (red) in the corresponding reachability digraph.}%
\label{fig:Bmap.9}%
\end{figure}

A road network can be represented by an arc-weighed digraph $D=(V,A, w:A\rightarrow \mathbb{R})$, where the set of vertices $V$ corresponds to road intersections and dead-ends, and an arc $(u,v)\in A$ represents a road segment connecting two corresponding vertices $u,v\in V$ in the specified direction. We define the weight $w(u,v)$ of an arc $(u,v)\in A$ as the length of the corresponding road segment, which also can be defined as energy or fuel consumption necessary to drive through the corresponding road segment. 
Notice that energy consumption can eventually be negative, e.g., in the case of regenerative braking, giving negative arc weights. 
An example of a digraph model for a road network in Birmingham, UK, is illustrated in Figure~\ref{fig:Bmap.9}, where the right diagram shows the road segment directions by using arcs for the central part of the left diagram. 
 
Given a road network digraph $D=(V,A, w:A\rightarrow \mathbb{R})$, a \emph{reachability digraph} $D'_r=(V',A'_r)$ can be defined as a simple digraph with $V' = V$ and arcs $(u,v)\in A'_r$ if and only if 
vertex $v$ can be reached from vertex $u$ by driving  in the road network within a given distance limit of $r$\,km.
Notice that there are no arc weights in $D'_r$ (if necessary, arc weights in $D'_r$ could be used too). 
The distance limit parameter $r$, defining the arcs of digraph $D'_r$, is called the \emph{reachability threshold} or \emph{radius}. 
A reachability digraph corresponding to a road network in Birmingham, UK, for the reachability radius $r=0.275$\,km is locally illustrated in the left diagram of Figure \ref{fig:Bmap.9}, where the size of the road map square is $825\,\mathrm{m}$\,$\times$\,$825\,\mathrm{m}$. 
Here, red dots indicate points of the road network reachable from the yellow point within the reachability radius of $0.275$\,km, i.e. they represent arcs from the yellow vertex to the red vertices in the corresponding $D'_{0.275}$.

Similarly to the reachability simple graph model described in \cite{GC2018}, 
the reachability digraph $D'_r$ models the situation where a driver becomes concerned about their low battery or fuel tank level and wishes to make a detour to stop at a refuelling facility reachable from their current location. One can assume that their vehicle will always have enough fuel or energy to reach a refuelling facility within a given reachability radius of $r$\,km. 
Thus, a reachability digraph does not need arc weights. 
Notice that assigning weights to the vertices of a reachability digraph can be used to model costs associated with opening and running the corresponding facilities. 

%%%%%%%%%%%%%%%%%%%
%%%%%%%%%%%%%%%
\subsection{Multiple dominating sets in digraphs}
\label{sec-domsets}

Given a subset of vertices $X \subseteq V$, a vertex $v \in V$ is said to be \emph{covered} by $X$ in $D$ if $v \in X$ or $N^-(v) \cap X \not= \emptyset $, i.e. $v$ is adjacent from a vertex in $X$. 
The set of vertices covered by $X$ in $D$ is denoted by $C(X) = X \cup \{ v \in V \mid N^-(v) \cap X \not= \emptyset \}$. If $C(X) = V$, then $X$ is called a \emph{dominating set} of $D$. 
Similarly, given an integer $k\ge 1$, a vertex $v \in V$ is said to be \emph{$k$-covered} by $X$ in $D$ if $v \in X$ or $|N^-(v) \cap X| \geq k$, i.e. $v$ is adjacent from at least $k$ vertices in $X$. 
The set of vertices $k$-covered by $X$ in $D$ is denoted by $C_k(X)$. 
Clearly, $C(X)=C_1(X)$.
If $C_k(X) = V$, then $X$ is a \emph{$k$-dominating set} of $D$.
In this case, the positive integer $k$ is called the \emph{multiplicity of domination} of $D$. 

A $k$-dominating set of the smallest cardinality in a digraph $D$ is called a \emph{minimum} $k$-dominating set. 
Its size is denoted by $\gamma_k(D)$ and called the \emph{$k$-domination number} of $D$.
In particular, $\gamma(D)=\gamma_1(D)$ is known as the \emph{domination number} of $D$.
A $k$-dominating set $X$ in $D$ is \emph{minimal} (by inclusion) if for any vertex $v\in X$, the set of vertices $X\setminus v$ is not $k$-dominating in $D$.
Given a digraph $D$, we are interested in finding a minimum $k$-dominating set in $D$, or else,
a reasonably small $k$-dominating set in $D$, possibly with some other features and properties for potential use in applications.
In particular, a $k$-dominating set $X$ in a reachability digraph $D'_r$ can indicate a set of locations in the corresponding road network such that, from any point without a refuelling facility in the network, a driver could reach at least $k$ of these $|X|$ locations within the reachability radius of $r$\,km to eventually refuel their vehicle.
More precisely, a point without a refuelling facility in the road network corresponds to a vertex $v$ of the associated road network digraph $D=(V,A, w:A\rightarrow \mathbb{R})$, such that $v$ is not in the $k$-dominating set $X$ of the corresponding reachability digraph $D'_r$. 

Notice that, if we use the above definitions and reachability model directly, with a $k$-dominating set defined via in-degrees and in-neighbours of the digraph (i.e., the requirement $|N^-(v) \cap X| \geq k$), a $k$-dominating set in the reachability digraph would represent a set of vertices where any location not in the set can be reached \emph{from} at least $k$ vertices of the set. This could be useful, e.g., for placement of fire stations or emergency-type facilities in general (e.g., defibrillators). 
However, in the case of refuelling facilities, we need to find a set of vertices such that, from any location without a facility in the road network, we can get \emph{to} at least $k$ vertices with facilities. Therefore, in this case, the direction of all arcs in the original reachability digraph must be reversed to indicate that a $k$-dominating set provides a set of destinations rather than origins. Alternatively, one could use out-degrees and out-neighbours, i.e., the inequality $|N^+(v) \cap X| \geq k$, to define a $k$-dominating set in the digraph in the opposite way. 

The problem of finding a minimum dominating set in a simple graph is one of the classic NP-hard problems \cite{GJ1979}. 
Moreover, this problem is known to be APX-hard (e.g., see \cite{RS1997}) and is not fixed parameter tractable \cite{DF1999} in general. 
The problem of finding a minimum $k$-dominating set in a simple graph is also known to be NP-hard \cite{LC2013}.
Since simple graphs can be considered as special cases of digraphs, corresponding computational complexity issues apply to digraphs.
Therefore, efficient and effective heuristics must be used to find reasonably good quality (small) $k$-dominating sets in digraphs.
To the best of our knowledge, no heuristics or efficient deterministic algorithms for finding such sets in digraphs are documented in the existing literature. 

The remainder of this paper is organized as follows. In Chapter~\ref{related}, we review some related results. 
A straightforward and two new experimentally optimized and tuned greedy heuristics are described in Chapter~\ref{heuristics}.
Also, in Chapter~\ref{heuristics}, we obtain an upper bound 
on the $k$-domination number $\gamma_k(D)$ of a digraph $D$, and describe 
a related new randomized heuristic, 
which can be considered as a flexible extension of the greedy heuristics.
In Chapter~\ref{Experiments}, computational experiments showing the efficiency and effectiveness of the proposed heuristics 
for two types of digraphs are presented, in particular, in comparison to an exact method, obtained using an integer linear programming (ILP)
formulation of the problem and a generic ILP solver (Gurobi \cite{Gurobi}). 
Our experiments also explore frontiers of tractability for the problem on very large random and road network digraphs. 
Finally, Chapter~\ref{Conclusion} provides a summary of the results, conclusions, and highlights opportunities for future research.

%%%%%%
%%%%%%%%%

\section{Related Work} % Literature review
\label{related}

Basic ideas for greedy heuristics for this kind of domination problems in simple graphs were previously discussed, e.g., in \cite{CG2021, GC2018, KL2004, P1991}.
In \cite{NS2020}, Nakkala and Singh considered greedy heuristics to search for small weight dominating sets in vertex-weighted digraphs.
Some basic theoretical results for the $k$-domination number of digraphs appeared in \cite{OBB2019}.
The new randomized algorithm of the current paper to search for $k$-dominating sets in digraphs is inspired by 
the ideas for simple graphs \cite{GC2018, GPZ2013} and also basic theoretical results for dominating sets in digraphs \cite{Lee1998}.
Notice that, as digraphs are a generalization of simple graphs, in general, 
concepts of dominating sets in digraphs are asymmetric and substantially different to simple graphs. 

%%%%%
In \cite{CG2021}, Corcoran and Gagarin proposed a greedy search heuristic for finding small $k$-dominating sets in simple graphs 
based on a problem formulation that highlights a certain structure of the $k$-domination problem.  
The proposed method was experimentally tested on a variety of reachability graphs corresponding to road networks of big cities (in the UK and abroad).
It is found that, for $k\ge 1$, their method is more accurate and outperforms the other methods, considered as benchmarks. 
The performance and improvements are also better for higher values of $k$. 
The authors also considered a generalization of the proposed greedy heuristic using beam search, which can be considered as a heuristic modification and a restriction of exhaustive search.

In \cite{GC2018}, Gagarin and Corcoran proposed a model for placement of refuelling facilities (for alternative fuel vehicles) 
based on multiple dominating sets in simple reachability graphs. 
Their experimental evaluation showed that, in this kind of graph, a randomized heuristic usually provides better results than a simple greedy heuristic, 
and its results are better suitable in applications. 
Nevertheless, the experimentally tuned and tested randomized heuristic used the simple greedy heuristic as a subroutine.
Two large road networks, Boston (USA) and Dublin (Ireland), were considered in computational experiments. 

Funke et al. \cite{FNS2015} model the problem of placing charging stations in road networks by considering digraphs, but mix terminology and notions of graphs and digraphs in their description. In digraph terms, they consider the problem as a ``shortest path" cover problem in a digraph $D=(V,A)$ corresponding to a road network. They try to find a small subset of vertices $L\subseteq V$ of the digraph such that every minimal (by inclusion) shortest path in $D$ exceeding the EV battery capacity has a charging station installed in a vertex of the set $L$. The problem is then reduced to the well-known hitting set problem, and an adaptation of the standard greedy approach is used, while developing some heuristic improvements to overcome limited computational resources for large-scale road networks.

%%%%%%

\section{Heuristic Algorithms and the Upper Bound}
\label{heuristics}

In this section, we describe three greedy heuristics, called Basic (Algorithm~\ref{BGH}), Deficiency Coverage (Algorithm~\ref{DCG}), and Two-Criteria (Algorithm~\ref{TCG}), as well as  
a randomized heuristic (Algorithm~\ref{RandAlg}) for finding small-size $k$-dominating sets in digraphs. 
The probabilistic method is used to obtain an upper bound on the $k$-domination number of a digraph,  
providing the randomized heuristic to search for $k$-dominating sets in the digraph.
The randomized heuristic is combined with and enhanced by the more effective search techniques of Two-Criteria Greedy, 
and can therefore be considered as a randomized extension of this. 
Usually, $k$-dominating sets initially found in digraphs by the main heuristics are not minimal (by inclusion). 
Therefore, another greedy heuristic (Algorithm~\ref{MDS}) is applied to an initially found $k$-dominating set at the end of each of the main heuristics to 
find a minimal $k$-dominating set in the input digraph.

%%%%%%%%%%%%%%%
\subsection{Greedy heuristics}
\label{GreedySection} 

A general basic greedy strategy for finding a small dominating set in a simple graph is well-studied in the literature \cite{AS1992,  P1991}. 
This starts with an empty set of vertices in the graph and iteratively extends a subset of vertices to a dominating set of the graph. 
In each iteration, the algorithm finds a vertex with the maximum amount of not covered neighbours with respect to a given subset of vertices 
and then adds the vertex to the subset. 
Repeating this process until all vertices are covered yields a dominating set of the graph. 
This basic greedy strategy can be generalized to the problem of finding $k$-dominating sets in simple graphs 
by iteratively checking for vertices that are not yet $k$-covered in the graph (e.g., see \cite{CG2021, GC2018}). 
In the case of digraphs, this basic greedy strategy can be adapted by iteratively checking the closed out-neighbourhood of each vertex of the digraph.
This provides the following heuristic, described in Algorithm \ref{BGH}.\\

\begin{algorithm}[H]
\caption{Basic Greedy}
\KwIn{A digraph $D=(V,A)$ and an integer $k\ge 1$.}
\KwOut{A minimal $k$-dominating set $Y$ of $D$.}
\BlankLine

\Begin{
	Initialize $X=\emptyset$\\
	\While{$C_k(X) \neq V$} {
		Find a vertex $\displaystyle v \in U = \argmax_{u \in V \setminus X} |N^+[u] \setminus C_k(X)|$\\
		Put $X = X \cup \{v\}$
	}
	Find a minimal $k$-dominating set $Y\subseteq X$\\  
	\mbox{}\ \ \ \ \ \ \ \ \ \ \ \ \ \ \ \ \ \ \ \ \ \ \ \ \ \ \ \ \ \ \ \ \ \tcc*[h]{e.g., see Algorithm~\ref{MDS} below}\\
	\Return $Y$
}
\label{BGH}
\end{algorithm}

%%%%%%%%%%%%%%%%
\vspace{5mm}

The $k$-dominating set $X\subseteq V$ of a digraph $D$ initially found in Algorithm~\ref{BGH} may be not minimal in $D$.
Clearly, we are interested in minimal $k$-dominating sets of $D$, and set $X$ may contain many subsets, of different cardinalities, which are $k$-dominating in D.
In general, finding a $k$-dominating subset of the smallest cardinality in a non-minimal $k$-dominating set $X$ of a digraph $D$ can be as challenging as finding a minimum $k$-dominating set in $D$, e.g., if we start with $X=V$. 

Given a $k$-dominating set $X$ of a digraph $D$, a vertex $v\in X$ is \emph{redundant} with respect to $X$ if $X\backslash \{v\}$ is also a $k$-dominating set in $D$, i.e. $C_k(X\backslash \{v\})=C_k(X)=V$.
It can be easily checked whether a vertex $v\in X$ is redundant with respect to $X$ in $D$ or not.
Denote by $R=\{v\in X\ |\ C_k(X\backslash \{v\})=V\}$ the set of all redundant vertices in $X$.
Suppose the initial $k$-dominating set $X$ found by Algorithm~\ref{BGH} in $D$ contains several redundant vertices, i.e. $|R|\ge 2$. Then, for any $v,w\in R$, $v\not=w$, although $X\backslash \{v\}$ and $X\backslash \{w\}$ are $k$-dominating in $D$, the set $X\backslash \{v,w\}$ may be not $k$-dominating in $D$, i.e. $C_k(X\backslash \{v,w\})\not=V$, e.g., when $v$ and $w$ have common out-neighbours in $D$. In other words, $w\in X$ may be not redundant with respect to $X\backslash \{v\}$, and $v$ may be not redundant with respect to $X\backslash \{w\}$ in $D$. So, removing one of the redundant vertices from $X$ may make some other redundant vertices with respect to $X$ not redundant with respect to a subset of $X$ in $D$.

Therefore, to try to remove from $X$ as many redundant vertices as possible, iteratively, one at a time, we use a heuristic greedy strategy that consists in removing redundant vertices of $X$ with lower impact on $k$-covered vertices of $D$ first. 
Specifically, to find a minimal $k$-dominating subset of $X$ in $D$, we prioritize removing redundant vertices of $X$ with a smaller number of out-neighbours in $V\backslash X$.
This is because removing a redundant vertex $v\in X$ from $X$ will lower the coverage of its $k$-covered out-neighbours in $N^+(v)\setminus X$, thus potentially forcing some redundant vertices of $X$ to become non-redundant with respect to $X\backslash \{v\}$ in $D$.
This is described in Algorithm \ref{MDS}.\\

\begin{algorithm}[H]
\caption{Minimal $k$-Dominating Subset}
\KwIn{A digraph $D=(V,A)$, an integer $k\ge 1$, a $k$-dominating set $X$ of $D$.}
\KwOut{A minimal $k$-dominating set $Y$ of $D$.}
\BlankLine

\Begin{
	\ForEach{$v \in X$} {
		Determine $x_v = |N^+(v)\setminus X|$ 
	}
	Initialize $Y = X$\\
	\While{$X \neq \emptyset$} {
		Find a vertex $\displaystyle v \in U = \argmin_{u \in X} x_u$\\
		\If{$C_k(Y \setminus \{v\}) = V$} {
			Put $Y = Y \setminus \{v\}$
		}
		Put $X = X \setminus \{v\}$
	}
	\Return $Y$
}
\label{MDS}
\end{algorithm}
%%%%%%%%%%%%%%%%%%%%%%%

\vspace{5mm}

To describe the next  {novel} 
greedy strategy, for each vertex $v$ of a digraph $D$, we assume that $v$ has $k$ \emph{slots} to be filled either by including $v$ into a $k$-dominating set or by having $k$ of its in-neighbours $N^-(v)$ in a $k$-dominating set of $D$. Therefore, in total, there are $k\cdot|V|$ slots to be filled by a $k$-dominating set in $D$. 
Given a subset $X\subseteq V$ of vertices of a digraph $D$ and an integer $k\ge 1$, the \emph{deficiency} of a vertex $v\in V\setminus X$ is defined as 
$$l_k(v,X)=\max\,\{0, k - |N^-(v) \cap X|\}.$$
The deficiency $l_k(v,X)$ indicates the level of coverage of the vertex by its in-neighbours in $X$ with respect to the required multiplicity of domination $k$.
If vertex $v$ is $k$-covered by $X$ in $D$, then $l_k(v,X)=0$.
Otherwise, $l_k(v,X) \ge 1$ indicates that $v$ requires $l_k(v,X)$ of its in-neighbours to be added to $X$ for $v$ to become $k$-covered from its in-neighbourhood $N^-(v)$ in $D$,
or, in other words, $l_k(v,X)$ of $v$'s own slots must be filled.
The \emph{slot coverage number} of a vertex $v\in V\backslash X$ with respect to $X$ in $D$ and an integer $k\ge 1$ is defined as $l_k(v,X)+|N^+(v) \setminus C_k(X)|$, i.e. the deficiency of $v$ plus the number of out-neighbours of $v$ not $k$-covered by $X$ in $D$.

Notice that, when $k\ge 2$, the Basic Greedy heuristic in Algorithm \ref{BGH} is not sensitive to the level of $k$-coverage of a vertex $v\in V\setminus X$ in each iteration:
$v$ itself is simply considered as either $k$-covered ($v\in C_k(X)$), implying $v\not\in N^+[v] \setminus C_k(X)$, or not $k$-covered ($v\not\in C_k(X)$), implying $v\in N^+[v] \setminus C_k(X)$. 
However, $k\ge 2$ implies that $v\in V\setminus X$ eventually needs several of its in-neighbours to be included into the set $X$ before it becomes $k$-covered from its in-neighbourhood $N^-(v)$ in $D$. 
On the other hand, if such a vertex $v$ is added to the set $X$, it would be $k$-covered even if none of its in-neighbours are in $X$. 
Therefore, it can be more beneficial to include into the set $X$ a vertex $v'\in V\backslash X$ with a smaller number of in-neighbours in $X$ rather than a vertex $v''\in V\backslash X$ that is almost $k$-covered by $X$ in $D$, 
even if such vertex $v''$ has more out-neighbours not $k$-covered by $X$ than $v'$ in $D$.

This gives rise to a more subtle greedy selection strategy,
accounting for how many slots inclusion of $v\in V\setminus X$ into a set $X\subseteq V$ would fill in $D$.
These heuristic ideas with the vertex slot coverage numbers are used in the Deficiency Coverage Greedy described in Algorithm \ref{DCG}. 
A minimal $k$-dominating set in Algorithm \ref{DCG} can also be found by using Algorithm~\ref{MDS}.\\

\begin{algorithm}[H]
\caption{Deficiency Coverage Greedy}
\KwIn{A digraph $D=(V,A)$, an integer $k\ge 1$.}
\KwOut{A minimal $k$-dominating set $Y$ of $D$.}
\BlankLine

\Begin{
	Initialize set $X=\emptyset$\\
	\While{$C_k(X) \neq V$} {
		Find a set $\displaystyle U = \argmax_{u \in V \setminus X}\,\,|N^+(u) \setminus C_k(X)|+ l_k(u,X)$\\
		Select $v\in U$ at random\\
		Put $X = X \cup \{v\}$
	}
	Find a minimal $k$-dominating set $Y\subseteq X$\\
	\mbox{}\ \ \ \ \ \ \ \ \ \ \ \ \ \ \ \ \ \ \ \ \ \ \ \ \ \ \ \ \ \ \ \ \ \ \ \ \ \ \ \ \ \ \ \ \ \ \ \  \ \ \ \tcc*[h]{using Algorithm~\ref{MDS}}\\
	\Return $Y$
}
\label{DCG}
\end{algorithm}

\vspace{5mm}

Another novel greedy strategy is based on refining the choice of the ``locally best" vertex in each iteration by using a secondary selection criterion to improve Algorithm \ref{DCG}.
Having several vertices with the same maximum slot coverage number in an iteration of Deficiency Coverage Greedy (one is to be included into a $k$-dominating set), the heuristic randomly selects one of them, uniformly at random.
Instead, when there are several such ``best candidates", it is possible to use the following greedy selection criterion.

For each vertex $v$ of a digraph $D=(V,A)$, its \emph{out-neighbourhood importance} factor is defined as
$$f_D(v)=\sum_{u\in N^+(v)} d^-(u),$$
i.e. $f_D(v)$ is the sum of all in-degrees of $v$'s out-neighbours in $D$.  
Now, we can assume that vertices of higher in-degree are more likely to be $k$-covered by their in-neighbours, rather than being included into a $k$-dominating set themselves.
Then, we can heuristically prioritize including vertices that have the out-neighbourhood importance factor higher.
This greedy selection criterion is considered whenever the slot coverage number $|N^+(u) \setminus C_k(X)|+ l_k(u,X)$ is maximized for several vertices $u\in V \setminus X$ in an iteration of Algorithm~\ref{DCG}.
The corresponding heuristic is described in Algorithm \ref{TCG} below.
A minimal $k$-dominating set in Algorithm \ref{TCG} is also found by using Algorithm~\ref{MDS}.\\

\begin{algorithm}[H]
\caption{Two-Criteria Greedy}
\KwIn{A digraph $D=(V,A)$, an integer $k\ge 1$.}
\KwOut{A minimal $k$-dominating set $Y$ of $D$.}
\BlankLine

\Begin{
	Initialize set $X=\emptyset$\\
	\While{$C_k(X) \neq V$} {
		Find a set $\displaystyle U = \argmax_{u \in V \setminus X}\,\,|N^+(u) \setminus C_k(X)|+ l_k(u,X)$\\
		Find a vertex $\displaystyle v \in U' = \argmax_{u \in U}\, f_D(u)$\\ 
		Put $X = X \cup \{v\}$
	}
	Find a minimal $k$-dominating set $Y\subseteq X$\\
	\Return $Y$
}
\label{TCG}
\end{algorithm}

\vspace{5mm}

For the worst-case complexity analysis, each of these greedy heuristics runs in $O(n^2+m)$ time. 
A more detailed analysis of our implementation of Two-Criteria Greedy shows that computing the out-neighbourhood importance factor $f_D(v)$ can be done in $d^+(v)$ steps for each vertex $v\in D$ before we start Algortihm~\ref{TCG}, which is $O(m)$ for all vertices of $D$ ($\sum_{v\in D}d^+(v)=m$).
Initializing the slot coverage numbers $l_k(v,\emptyset) + |N^+(v) \setminus C_k(\emptyset)|$ for all $v\in D$ can be done in $O(n)$ time. 
Each iteration of the while loop in Algortihm~\ref{TCG} takes $O(n)$ time to find set $U$ and then another $O(n)$ time to find a vertex from $U$ with the largest out-neighbourhood importance factor. 
This gives $O(n^2)$ time over all iterations. 
Updating the coverage provided by the set $X:=X\cup \{v\}$ and the uncovered out-neighbourhoods of the vertices not yet included in $X$ in each iteration requires at most $d^-(v)$ adjustments by $1$ unit for the in-neighbours of the newly selected vertex $v$, giving at most $\sum_{v\in D}d^-(v)=m$ over all iterations.
Similarly, at most $d^+(v)$ out-neighbours of $v$ need to have their level of coverage adjusted by $1$ unit, giving at most $\sum_{v\in D}d^+(v)=m$ in the worst case.
Also, in each iteration, an out-neighbour $v'$ of $v$ can become $k$-covered by $X$, which requires an update for the coverage power of its in-neighbours at most once over all iterations, giving at most $\sum_{v'\in D}d^-(v')=m$ steps overall. 
Since each vertex of $D$ is $k$-covered by its in-neighbours or included in $X$ (possibly both) once during the algorithm run, we have $O(m)$ time complexity over all iterations for updates and checks of the $k$-coverage. Thus, overall, Algortihm~\ref{TCG} runs in $O(n^2+m)$ time.

%%%%%%%%%%%%%%%

\subsection{Randomized heuristic}
\label{RandomizedSection}

The new randomized heuristic is developed by using a basic form of the probabilistic method (e.g., see \cite{AS1992, GPZ2013, Lee1998}),
which is enhanced by greedy heuristic ideas from the previous section.
Alternatively, this randomized heuristic can be considered as an extension and a generalization of Two-Criteria Greedy, where the set $X$ is initialized to be a random subset of vertices in the digraph, instead of an empty set.
This allows Two-Criteria Greedy to become more flexible, instead of just following its rather rigid greedy selection criteria from scratch.

%%%%%%
\subsubsection{Using the probabilistic method}

The probabilistic method, which is used here as an analytic tool, can be sketched as follows. 
Suppose we need to estimate the size of a minimum dominating set in a simple graph $G=(V,E)$.
First, using a probability $p\in \left[0,1\right]$, which is supposed to be optimized, we decide whether to include each vertex $v$ of $G$ into a set $X$ of vertices or not.
Then, considering the set $B$ of all vertices not covered by $X$ in $G$, i.e. $B=V\backslash C(X)$, 
we have a dominating set $X\cup B$ of $G$.
Since probabilities of both events, of a vertex being included in $X$ and of a vertex being included in $B$, can be computed explicitly, 
the expected cardinality $\mathbb{E}(|X|) + \mathbb{E}(|B|)$ of the dominating set $X\cup B$ in $G$ can be computed as well. 
As there is at least one dominating set of cardinality at most $\mathbb{E}(|X|) + \mathbb{E}(|B|)$, this yields an upper bound on the cardinality of minimum dominating sets in $G$. 
This approach has been generalized by Gagarin et al. \cite{GPZ2013} to $k$-dominating sets in simple graphs, giving the following upper bound and the corresponding randomized heuristic.

\begin{theorem}[\cite{GPZ2013,Zve2}]
\label{ThmClassic}
	Given a simple graph $G$ of minimum vertex degree $\delta=\delta(G)$ and an integer $k$, $1\le k\le \delta$, we have 
	$$\displaystyle \gamma_k(G) \leq \left( 1 - \frac{\delta^\prime}{{\delta \choose k-1}^{1/\delta^\prime}\cdot(1 + \delta^\prime)^{1 + 1/\delta^\prime}} \right) n,$$
	where $\delta^\prime = \delta -k +1$.
\end{theorem}

The optimized probability $\displaystyle p=1 - \frac{1}{\sqrt[\delta^\prime]{{\delta \choose k-1}(1 + \delta^\prime)}}$ and 
the corresponding randomized algorithm follow from the proof of Theorem~\ref{ThmClassic} (see \cite{GPZ2013} for details). 
The algorithm finds a $k$-dominating set in a simple graph $G$. 
Some experimental results with this algorithm are presented in \cite{GC2018}.
When $k=1$, this algorithm can be easily derandomized by using Basic Greedy (e.g., see \cite{AS1992}).
However, already for $k=2$, the task of derandomizing this heuristic becomes much more challenging.
For example, see the results of Harant and Henning \cite{HH2008} for a similar concept of double domination in simple graphs.
Also, notice that, considering a simple graph $G$ as the corresponding digraph $D$, we have $\delta(G)=\delta^{+}(D)=\delta^{-}(D)$, which is not the case in general digraphs. 

%%%%%%%%%%%%
Hence, our randomized heuristic to find small $k$-dominating sets in digraphs is based on the following application of the probabilistic method. This can be considered as a generalization of the corresponding result for $k=1$ by Lee \cite{Lee1998}.

\begin{theorem}
\label{ProbBound}
	Given a digraph $D=(V,A)$ with its minimum in-degree $\delta^{-}=\delta^{-}(D)$ and an integer $k$, $1\le k\le \delta^{-}$, we have 
	$$\displaystyle \gamma_k(D) \leq \left( 1 - \frac{\delta^{-\prime}}{{\delta^{-} \choose k-1}^{1/\delta^{-\prime}}\cdot(1 + \delta^{-\prime})^{1 + 1/\delta^{-\prime}}} \right) n,$$
	where $\delta^{-\prime} = \delta^{-} -k +1$.
\end{theorem}

\proof First, for each vertex $v \in V$, select $\delta^-$ of its in-neighbours to form a subset $N_{\delta^-}(v)$ of the $v$'s in-neighbourhood $N^{-}(v)$ in $D$. 
Then, form an initial subset of vertices $X\subseteq V$ by deciding with probability $p$ independently for each vertex $v$ of $D$ whether $v$ is included into $X$ or not.
The remaining vertices of $V \setminus X$ can be partitioned into subsets $B_m$,\ \,$m=0,1,\ldots,\delta^-$,\ \,as follows:
$$B_m = \{v \in V \setminus X\,\ \mid\,\  N_{\delta^-}(v) \cap X = m\}.$$
Now, since every vertex not $k$-covered by $X$ in $D$ has at most $k-1$ of its in-neighbours in $X$,
the set of not $k$-covered by $X$ in $D$ vertices is a subset of $\displaystyle B=\bigcup_{m=0}^{k-1} B_m \subseteq V \setminus X$, and $X\cup B$ is $k$-dominating in $D$.

Therefore, we have 
$$\gamma_k(D) \leq \mathbb{E}(|X|) + \mathbb{E}(|B|) \leq \mathbb{E}(|X|) + \sum_{m = 0}^{k-1} \mathbb{E}(|B_m|).$$
Then, $\mathbb{E}(|X|)=pn$, and, for each of the subsets $B_m$,\ \,$m=0,1,\ldots,k-1$,
\begin{align*}
\mathbb{E}(|B_m|) & = \sum_{v \in V} \mathbb{P}(v \in B_m) \\
				  & = \sum_{v \in V} (1-p )\genfrac(){0pt}{0}{\delta^-}{m} p^m (1-p)^{\delta^- - m} \\
				  & = n \genfrac(){0pt}{0}{\delta^-}{m} p^m (1-p)^{\delta^- - m + 1} \\
				  & \leq n \genfrac(){0pt}{0}{\delta^- - m}{\delta^- - k + 1} \genfrac(){0pt}{0}{\delta^-}{m} p^m (1-p)^{\delta^- - m + 1} \\
				  & = n \genfrac(){0pt}{0}{\delta^-}{k-1} \genfrac(){0pt}{0}{k-1}{m} p^m (1-p)^{\delta^- - m + 1}\ \ . 
\end{align*}
From this,
\begin{align*}
\gamma_k(G) & \leq np + \sum_{m = 0}^{k-1} n \genfrac(){0pt}{0}{\delta^-}{k-1} \genfrac(){0pt}{0}{k-1}{m} p^m (1-p)^{\delta^- - m + 1} \\
			& = np + n \genfrac(){0pt}{0}{\delta^-}{k-1} (1-p)^{\delta^- - k + 2} \sum_{m = 0}^{k-1} \genfrac(){0pt}{0}{k-1}{m} p^m (1-p)^{k - 1 - m} \\
			& = np + n \genfrac(){0pt}{0}{\delta^-}{k-1} (1-p)^{\delta^- - k + 2}\cdot (p+1-p)^{k-1}\\
			& = np + n \genfrac(){0pt}{0}{\delta^-}{k-1} (1-p)^{\delta^{-'} + 1}\ \ . 
\end{align*}
Minimizing the last expression as a function of $p$, more specifically,
$$\displaystyle f(p)=n \left(p + \genfrac(){0pt}{0}{\delta^-}{k-1} (1-p)^{\delta^{-'} + 1}\right),$$
for $p\in [0,1]$, we have 
$$\displaystyle p=1 - \frac{1}{\sqrt[\delta^{-\prime}]{{\delta^{-} \choose k-1}(1 + \delta^{-\prime})}}$$ 
and 
$$\gamma_k(G) \leq n \left( 1 - \frac{\delta^{-\prime}}{{\delta^{-} \choose k-1}^{1/\delta^{-\prime}}\cdot(1 + \delta^{-\prime})^{1 + 1/\delta^{-\prime}}} \right).$$
\qed

%%%%%%%
%%%%%%%%%%
\subsubsection{Combining and tuning the randomized and greedy heuristics}

We can apply the probabilistic method ideas used in the proof of Theorem~\ref{ProbBound} to find small $k$-dominating sets in digraphs, satisfying the upper bound of Theorem~\ref{ProbBound} and eventually improving the outcomes of greedy heuristics from Section~\ref{GreedySection}.
Notice that the value of probability $p$ computed 
in the proof of Theorem~\ref{ProbBound} is optimized for digraphs in general, which may be not subtle enough for some particular digraphs.
An initially selected random subset $X$ of vertices of a digraph $D$ can be extended and then adjusted to a minimal $k$-dominating set in $D$.
This can be done  much more effectively than in the proof of Theorem~\ref{ProbBound} by using heuristic ideas, e.g., from the greedy algorithms.
The following novel randomized algorithm (Algorithm~\ref{RandAlg}) can be applied several times to the same digraph, 
using the same probability $p$ or different values for the probability $p$, 
to obtain different initial subsets $X$ of vertices in $D$, of different cardinalities. 
A greedy heuristic can use this flexibility and an initial non-empty random subset $X$ of vertices in $D$ (instead of an empty set) before starting its greedy iteration to find a small $k$-dominating set in $D$. A pseudocode for this general randomized heuristic is presented in Algorithm~\ref{RandAlg}.\\

\begin{algorithm}[H]
\label{RandAlg}
\caption{Randomized Heuristic}
\KwIn{A digraph $D=(V,A)$, an integer $k\ge 1$.}
\KwOut{A minimal $k$-dominating set $Y$ of $D$.}
\BlankLine

\Begin{
	Initialize set $X = \emptyset$;\\
	Compute an inclusion probability $p$;\\
	\ForEach{$v \in V$} {
		Using probability $p$, decide whether to include $v$ into $X$ or not;
	}
	Use a greedy heuristic to add vertices into the set $X$ to obtain a $k$-dominating set $X'$ in $D$;\\
	Find a minimal $k$-dominating set $Y\subseteq X'$;\\
	\Return $Y$
}
\end{algorithm}

For the computational experiments presented in Section~\ref{Experiments}, Two-Criteria Greedy (Algorithm~\ref{TCG}) was used to complete an initial random set $X$ in Algorithm~\ref{RandAlg} to a  $k$-dominating set $X'$ of a digraph $D$. 
A minimal $k$-dominating set in Algorithm~\ref{RandAlg} was found by using the greedy heuristic of Algortihm~\ref{MDS}.

Our computational experiments show that performance of Algorithm~\ref{RandAlg} strongly depends on probability $p$. 
In general, following the application of probabilistic method in Theorem~\ref{ProbBound}, the probability $p$ in Algorithm~\ref{RandAlg} is optimized when 
$$p=1 - \frac{1}{\sqrt[\delta^{-\prime}]{{\delta^{-} \choose k-1}(1 + \delta^{-\prime})}},$$
where $\delta^{-\prime} = \delta^{-} -k +1$. In other words, assuming $k$ is fixed, $p$ can be considered as a function of the minimum in-degree $\delta^{-}$ of a  digraph $D$. 
However, since this probability is optimized in general digraphs, its value is too high in some particular classes of digraphs, and it gives a very large initial subset of vertices $X$ of the digraph $D$ (to satisfy the upper bound of Theorem~\ref{ProbBound}). 
Therefore, as shown by the corresponding computational experiments for simple graphs in \cite{GC2018}, it makes sense to consider $p$ as a function of the same form as in Theorem~\ref{ProbBound}, but of some other vertex in-degree parameters of the digraph $D$, e.g. the average in-degree $d^-_{\mathrm{ave}}$ or the median in-degree $d^-_{\mathrm{med}}$ of $D$, instead of $\delta^{-}$.
More generally, for a fixed value of $k$, $1\le k\le \delta^{-}$, we can heuristically estimate and adjust the value of probability $p$ by considering it as a function $p=p(x)$ of the form
\begin{equation}
p(x)=1 - \frac{1}{\sqrt[x-k+1]{{\lfloor x\rfloor \choose k-1}(x-k+2)}},
\end{equation}
where $x$ is a vertex in-degree parameter, $x\in [\delta^{-}, \Delta^{-}]$,\ \ $\delta^{-}=\delta^{-}(D)$, $\Delta^{-}=\Delta^{-}(D)$.
Then, for example, using $x\in \{\delta^{-},d^-_{\mathrm{ave}},d^-_{\mathrm{med}},\Delta^{-}\}$ and some other values of $x$ from the interval $[\delta^{-}, \Delta^{-}]$ allows us to experiment with Algorithm~\ref{RandAlg} and to tune it to obtain better results. 

For the worst-case complexity analysis of this randomized heuristic, it takes an $O(n)$ time to find an initial random set $X$ in Algorithm~\ref{RandAlg} when probability $p$ is known. However, computing probability $p$ in Algorithm~\ref{RandAlg} involves computing the binomial coefficient, which can be done in $O(n^2)$ time. Since Two-Criteria Greedy (Algorithm~\ref{TCG}) is used to complete an initial random set $X$ in Algorithm~\ref{RandAlg} to a  $k$-dominating set $X'$ of a digraph $D$, and then
a minimal $k$-dominating set is found by using the greedy heuristic of Algortihm~\ref{MDS}, the overall time complexity of Algorithm~\ref{RandAlg} is still $O(n^2+m)$. 

%%%%%%%%%%
\section{Computational Experiments}
\label{Experiments}

Using the basic and new greedy and randomized heuristics described in Section~\ref{heuristics}, 
a number of experiments were run to test effectiveness and efficiency of the heuristics.
Results of these computational experiments are presented in this section. 
We use abbreviations BG, DCG, and TCG for the Basic Greedy, Deficiency Coverage Greedy, and Two-Criteria Greedy heuristics, respectively. 
We write ILP for results returned by the generic integer linear programming solver Gurobi \cite{Gurobi}.

Two types of digraphs were considered in this computational study. 
The first type are random digraphs generated according to the so-called Erd\H{o}s-R\'enyi model. In this random digraph model, 
every arc is included (or not) into the digraph independently with some fixed probability $p$ (or, respectively, $1-p$).  
It is important to note that, in contrast to simple graphs, every pair of vertices here corresponds to two possible arcs directed in opposite ways. 
These two arcs are considered to be distinct and therefore included or not independently of each other, like two different edges in a simple graph. 
This means that any given pair of vertices has probability $2p(1-p)$ of being connected by a single arc in either direction, 
and probability $p^2$ of being connected in both directions.
The probability of two vertices being adjacent is thus $p(2-p)$, as opposed to the corresponding simple graph random model, 
where this probability is simply $p$. 

The second type of digraphs are reachability digraphs associated with actual road networks. They are described and formally defined in Section~\ref{sec:reachability}.
In a reachability digraph, an arc $(u,v)$ indicates that there is a way of travelling from vertex $u$ to vertex $v$ within a certain pre-established distance of $r$\,km in the corresponding road network. 
Since most streets support two-way traffic, many vertex pairs in such a digraph will either have both corresponding arcs included or be non-adjacent at all.
In comparison to the Erd\H{o}s-R\'enyi digraphs, this makes reachability digraphs more similar to simple graphs.
However, there is still a significant amount of one-way streets, 
as well as roads like motorways, where both directions are effectively separate streets that often connect to different exit roads.
Therefore, it is still possible and indeed relatively common for an arc $(u,v)$ to be included, whilst its counterpart $(v,u)$ is not. 

For each Erd\H{o}s-R\'enyi and reachability digraph, four different types of $k$-domination were considered. 
Dominating sets, i.e. $1$-dominating sets, were considered alongside $k$-dominating sets for $k=2, 4$, and $8$.
Whenever a randomized algorithm performed a random selection of an initial vertex set,
it was run $10$ times, and the best (out of $10$) result was then chosen and reported. 
The heuristics for these experiments were implemented using C++ and executed on a desktop PC with a 3.00 GHz Intel Core i5 processor and 16 GB of RAM, running Windows 10 Education, version 21H2.

%%%%%
\subsection{ILP formulation and experiments with a deterministic ILP solver}
For both digraph types, a number of smaller size problem instances were considered   
in order to be able to compare some of the heuristic results to the corresponding optimal solutions. 
These (deterministic) optimal solutions were obtained using Gurobi 10.0.1 to solve the following 
integer linear programming (ILP) formulation of the problems.

A binary decision variable $x_i$ is associated with each vertex $v_i$ of a digraph $D$, $i = 1,\dots,n$, where $n=|V_D|$. 
These decision variables indicate wether a vertex is included in a $k$-dominating set or not: 
$x_i = 0$ means vertex $v_i$ is not in the set, while $x_i = 1$ means $v_i$ is in the set.
Therefore, the sum of all $x_i$'s equals to the size of the set. 
Then, to find a minimum $k$-dominating set in $D$, one can solve the following ILP problem:
\begin{equation}
\begin{array}{ll@{}lrl}
\text{minimize}  	& z(x_1,x_2,\ldots,x_n) = \displaystyle \sum \limits_{i=1}^{n} x_i 		& \\[7mm]

\text{subject to:}	& kx_i + \displaystyle \sum \limits_{v_{j} \in N^-(v_i)} x_{j} \geq k,  	&i=1,\dots,n & \label{constraint}\\[7mm]

					& x_{i} \in \{0,1\} 													&i=1,\dots,n & \\[7mm]
\end{array}
\end{equation}
Constraints (\ref{constraint}) ensure that each vertex $v_i$, $i=1,\dots,n$, is $k$-dominated by the set.
If $v_i$ itself is included in the set, then $kx_i = k$, which is already enough to satisfy the corresponding constraint. 
Otherwise, the constraint is only satisfied if at least $k$ of $v_i$'s in-neighbours in $D$ are in the set.

For higher values of $k$ and some of the small digraphs, the ILP experiments to find the exact value of $\gamma_k(D)$ required unreasonably high amount of computing time.
Therefore, a CPU time limit of $30$ minutes was imposed on each run of the ILP solver ($24$ hours for some of the runs in extended experiments). 
If the CPU time limit was reached by Gurobi without returning an optimal solution, the final best feasible solution found by Gurobi was considered as a heuristic solution instead, so that it could be compared to other heuristic solutions. 
Naturally, it was impractical to run experiments with a deterministic ILP solver for large size problem instances, 
which was clearly indicated in the experiments.

%%%%%%%%
\subsection{Erd\H{o}s-R\'enyi digraphs}
The algorithms were tested on a number of Erd\H{o}s-R\'enyi digraphs. 
To generate each such digraph, the arc inclusion probability of $p = 0.1$ was used, 
which was motivated by having more challenging problem instances in sparse digraphs (e.g., see \cite{Wendy1} for simple graphs) 
while keeping in- and out-neighbourhoods reasonably large and the digraphs reasonably connected. 
BG, DCG, and TCG were used alongside the randomized heuristic based on the minimum and average vertex in-degrees.
Since the Erd\H{o}s-R\'enyi random digraph model generates digraphs with very similar or almost equal average and median in-degrees, 
the randomized algorithms return near-identical results when using one of these two parameters. 
Therefore, the randomized algorithm based on the median vertex in-degree was not considered in these experiments.
Randomized algorithms using the maximum in-degree, as well as the first and third quartile values of the in-degree sequence were also considered, 
but turned out to be less effective than those using the minimum and average in-degrees.

The smaller scale experiments for this type of digraph were run on digraphs with $100$, $150$, and $200$ vertices. 
The corresponding results are presented in Table~\ref{ER100}, 
where the numbers in parentheses indicate the best (heuristic) results found by the generic ILP solver within the given time limit (30\,min), 
and the best (out of three) results returned by the greedy heuristics are highlighted. 
The time limit of 30\,min was chosen to have the results comparable in time with our greedy heuristics in general: the greedy heuristics use less than 30\,min in all of our experiments.

%%%%%%%%%%%%
\begin{table}[H]
	\centering
	\begin{tabular}{| c | c | r | r | r | r | r | r | r | r | }
	\hline
\multirow{2}{*}{$n$} &	\multirow{2}{*}{$k$} & \multicolumn{2}{c|}{ILP}          & \multicolumn{2}{c|}{BG}        & \multicolumn{2}{c|}{DCG}        & \multicolumn{2}{c|}{TCG}        \\ \cline{3-10} 
&					   & \multicolumn{1}{c|}{Size} & Time\,(s) & \multicolumn{1}{c|}{Size} & Time\,(s)   & \multicolumn{1}{c|}{Size} & Time\,(s)   & \multicolumn{1}{c|}{Size} & Time\,(s)  \\ 
\hline 
\multirow{4}{*}{$100$} &	1                  & \multicolumn{1}{r|}{12}   & 0.2   & \multicolumn{1}{r|}{14}   & 0.00057 & \multicolumn{1}{r|}{\cellcolor{orange!25} 12}   & 0.00059 & \multicolumn{1}{r|}{\cellcolor{orange!25} 12}   & 0.00055 \\ \cline{2-10}  
&	2                  & \multicolumn{1}{r|}{21}   & 1.98   & \multicolumn{1}{r|}{\cellcolor{orange!25} 24}   & 0.00067 & \multicolumn{1}{r|}{\cellcolor{orange!25} 24}   & 0.00073 & \multicolumn{1}{r|}{\cellcolor{orange!25} 24}   & 0.00069  \\ \cline{2-10} 
&	4                  & \multicolumn{1}{r|}{36}   & 5.72  & \multicolumn{1}{r|}{45}   & 0.0009 & \multicolumn{1}{r|}{\cellcolor{orange!25} 38}   & 0.00091 & \multicolumn{1}{r|}{40}   & 0.00084 \\ \cline{2-10} 
&	8                  & \multicolumn{1}{r|}{64}   & 1.46   & \multicolumn{1}{r|}{71}   & 0.0011  & \multicolumn{1}{r|}{\cellcolor{orange!25} 66}   & 0.0012  & \multicolumn{1}{r|}{\cellcolor{orange!25} 66}   & 0.0011 \\ 
\hline 
\multirow{4}{*}{$150$} &	1                  & \multicolumn{1}{r|}{14}   & 10.45    & \multicolumn{1}{r|}{\cellcolor{orange!25} 15}   & 0.00087 & \multicolumn{1}{r|}{\cellcolor{orange!25} 15}   & 0.00093 & \multicolumn{1}{r|}{\cellcolor{orange!25} 15}   & 0.00089  \\ \cline{2-10}  
&	2                  & \multicolumn{1}{r|}{22}   & 49.46   & \multicolumn{1}{r|}{25}   & 0.001  & \multicolumn{1}{r|}{25}   & 0.0013  & \multicolumn{1}{r|}{\cellcolor{orange!25} 24}   & 0.001    \\ \cline{2-10}  
&	4                  & \multicolumn{1}{r|}{(39)}   & 1,800 & \multicolumn{1}{r|}{\cellcolor{orange!25} 42}   & 0.0013  & \multicolumn{1}{r|}{\cellcolor{orange!25} 42}   & 0.0014  & \multicolumn{1}{r|}{43}   & 0.0013  \\ \cline{2-10}   
&	8                  & \multicolumn{1}{r|}{(71)}   & 1,800 & \multicolumn{1}{r|}{76}   & 0.0024  & \multicolumn{1}{r|}{\cellcolor{orange!25} 74}   & 0.005  & \multicolumn{1}{r|}{77}   & 0.0018  \\ 
\hline 
 \multirow{4}{*}{$200$}  & $1$                  & \multicolumn{1}{r|}{15}   & 691 & \multicolumn{1}{r|}{19}   & 0.0014 & \multicolumn{1}{r|}{17}   & 0.0014 & \multicolumn{1}{r|}{\cellcolor{orange!25} 16}   & 0.0014 \\ \cline{2-10}
 & $2$                  & \multicolumn{1}{r|}{(25)}   & 1,800   & \multicolumn{1}{r|}{28}   & 0.0015 & \multicolumn{1}{r|}{\cellcolor{orange!25} 27}   & 0.0016 & \multicolumn{1}{r|}{\cellcolor{orange!25} 27}   & 0.0015 \\ \cline{2-10}
 & $4$                  & \multicolumn{1}{r|}{(44)}   & 1,800   & \multicolumn{1}{r|}{47}   & 0.003    & \multicolumn{1}{r|}{47}   & 0.0021 & \multicolumn{1}{r|}{\cellcolor{orange!25} 46}   & 0.0019 \\ \cline{2-10}
 & $8$                  & \multicolumn{1}{r|}{(78)}   & 1,800    & \multicolumn{1}{r|}{85}   & 0.0027 & \multicolumn{1}{r|}{\cellcolor{orange!25} 83}   & 0.0041 & \multicolumn{1}{r|}{\cellcolor{orange!25} 83}   & 0.0071  \\ \hline 
	\end{tabular} 
	\caption{ILP solver and greedy heuristics results for $k$-dominating sets in small Erd\H{o}s-R\'enyi digraphs.}
	\label{ER100}
\end{table}

Table~\ref{ER100} shows that the proposed greedy heuristics solve the small size problem instances in milliseconds, while the solution quality is comparable to the exact or heuristic ILP solutions, after running the generic ILP solver on these small size instances for a much longer time. The best (out of three) greedy solution quality matches the optimal solution in one case, and is within $14.3\%$ of the optimum in all the other cases.
The problem instances that Gurobi was not able to solve to the optimum in $30$ minutes (size numbers in parentheses in Table~\ref{ER100}), were not solved by the deterministic ILP solver to the optimum even in $24$ hours. 

%%%%%%%%%%%%%%%%

The main set of experiments were run on large digraphs with 25,000, 50,000, 75,000, and 100,000 vertices.
An attempt was also made to run Gurobi on the Erd\H{o}s-R\'enyi digraph with 25,000 vertices, but the generic solver was able to find only initial (heuristic) solutions of size $68, 89, 125$, and $185$ vertices for $k=1,2,4,$ and $8$, respectively, before running out of memory. 
An experiment with our greedy heuristics on a 125,000-vertex digraph was attempted, but the computer did not have sufficient memory and was not able to execute it. 
Results for the large-scale experiments returned by the greedy heuristics on the other four large digraphs are shown in Table~\ref{ER100000},
where the best (out of three) results returned by the greedy heuristics are highlighted.
Using the randomized heuristics on Erd\H{o}s-R\'enyi digraphs did not improve the results of BG, DCG, and TCG, only matched some of them. 
Therefore, the corresponding results are omitted from the tables.

\begin{table}[H]
	\centering
	\begin{tabular}{| c | c | r | r | r | r | r | r |}
  \hline
  \multirow{2}{*}{$n$}  &  \multirow{2}{*}{$k$} & \multicolumn{2}{c|}{BG}      & \multicolumn{2}{c|}{DCG}    & \multicolumn{2}{c |}{TCG}    \\ \cline{3-8} 
  &                    & \multicolumn{1}{c|}{\small{Size}} & \small{Time\,(s)} & \multicolumn{1}{c|}{\small{Size}} & \small{Time\,(s)} & \multicolumn{1}{c|}{\small{Size}} & \small{Time\,(s)} \\  
  \hline 
 \multirow{4}{*}{25,000} &	1                & \multicolumn{1}{r|}{\cellcolor{orange!25} 47}   & 3.97   & \multicolumn{1}{r|}{\cellcolor{orange!25} 47}   & 3.91   & \multicolumn{1}{r|}{\cellcolor{orange!25} 47}   & 4.9 \\  \cline{2-8} 
&	2                  & \multicolumn{1}{r|}{\cellcolor{orange!25} 65}   & 4.66   & \multicolumn{1}{r|}{66}   & 4.32   & \multicolumn{1}{r|}{\cellcolor{orange!25} 65}   & 5.33 \\  \cline{2-8} 
&	4                  & \multicolumn{1}{r|}{97}   & 5.1   & \multicolumn{1}{r|}{\cellcolor{orange!25} 95}   & 4.85   & \multicolumn{1}{r|}{97}   & 6.37 \\  \cline{2-8} 
&	8                  & \multicolumn{1}{r|}{\cellcolor{orange!25} 152}  & 5.49   & \multicolumn{1}{r|}{154}  & 5.13  & \multicolumn{1}{r|}{153}  & 6.01 \\ 
\hline 
 \multirow{4}{*}{50,000} & $1$                  & \multicolumn{1}{r|}{52}   & 21.95   & \multicolumn{1}{r|}{\cellcolor{orange!25} 51}   & 22.34   & \multicolumn{1}{r|}{52}   & 25.47   \\ \cline{2-8} 
 & $2$                  & \multicolumn{1}{r|}{73}   & 23.27   & \multicolumn{1}{r|}{\cellcolor{orange!25} 72}   & 21.68   & \multicolumn{1}{r|}{\cellcolor{orange!25} 72}   & 25.95   \\ \cline{2-8} 
 & $4$                  & \multicolumn{1}{r|}{105}  & 23.21   & \multicolumn{1}{r|}{\cellcolor{orange!25} 104}  & 21.65   & \multicolumn{1}{r|}{105}  & 26.29   \\ \cline{2-8} 
 & $8$                  & \multicolumn{1}{r|}{165}  & 23.72   & \multicolumn{1}{r|}{164}  & 23.16   & \multicolumn{1}{r|}{\cellcolor{orange!25} 163}  & 25.9    \\ 
  \hline 
 \multirow{4}{*}{75,000}  & 	1                  & \multicolumn{1}{r|}{\cellcolor{orange!25} 55}   & 251.3   & \multicolumn{1}{r|}{\cellcolor{orange!25} 55}   & 251.5   & \multicolumn{1}{r|}{\cellcolor{orange!25} 55}   & 255.5  \\ \cline{2-8} 
&	2                  & \multicolumn{1}{r|}{\cellcolor{orange!25} 75}   & 292     & \multicolumn{1}{r|}{76}   & 304.8   & \multicolumn{1}{r|}{76}   & 341.9  \\ \cline{2-8} 
&	4                  & \multicolumn{1}{r|}{110}  & 307.4   & \multicolumn{1}{r|}{110}  & 288.3   & \multicolumn{1}{r|}{\cellcolor{orange!25} 109}  & 285.1  \\ \cline{2-8} 
&	8                  & \multicolumn{1}{r|}{169}  & 335.8   & \multicolumn{1}{r|}{169}  & 310.9   & \multicolumn{1}{r|}{\cellcolor{orange!25} 168}  & 317.8  \\ 
\hline
\multirow{4}{*}{100,000} & $1$                  & \multicolumn{1}{r|}{58}   & 1,654    & \multicolumn{1}{r|}{58}   & 1,392    & \multicolumn{1}{r|}{\cellcolor{orange!25} 57}   & 1,407    \\ \cline{2-8} 
 & $2$                  & \multicolumn{1}{r|}{\cellcolor{orange!25} 78}   & 1,470    & \multicolumn{1}{r|}{\cellcolor{orange!25} 78}   & 1,422    & \multicolumn{1}{r|}{\cellcolor{orange!25} 78}   & 1,453    \\ \cline{2-8} 
 & $4$                  & \multicolumn{1}{r|}{113}  & 1,521    & \multicolumn{1}{r|}{113}  & 1,576    & \multicolumn{1}{r|}{\cellcolor{orange!25} 112}  & 1,587    \\ \cline{2-8} 
 & $8$                  & \multicolumn{1}{r|}{174}  & 1,574    & \multicolumn{1}{r|}{173}  & 1,646    & \multicolumn{1}{r|}{\cellcolor{orange!25} 172}  & 1,598   \\  \hline
  \end{tabular}
	\caption{Greedy heuristics results for $k$-dominating sets in large Erd\H{o}s--R\'enyi digraphs.}
	\label{ER100000}
\end{table}

Comparing the greedy solvers to each other, Tables~\ref{ER100} and \ref{ER100000} show that, for most of the problem instances, DCG and TCG find better quality greedy solutions. However, BG finds the best (out of three) greedy solution in 2 cases and matches the best quality greedy solution in 7 of the other cases, out of the total 28 problem instances. 
Therefore, in the case of Erd\H{o}s-R\'enyi digraphs, BG cannot be ignored.
The runtime of each of the greedy solvers is within a reasonable time limit for all of the problem instances (less than $30$\,min for a digraph on 100,000 vertices).

%%%%%%%%%%%%%%%%%%%%%%%%%%%%%%%%%
\subsection{Road networks of Birmingham and the West Midlands conurbation}

The road network digraphs were obtained from OpenStreetMap \cite{OSM} using the OSMnx library in the programming language Python.
The distance between a pair of vertices of a road network digraph was measured as the length of a shortest path between the vertices. 
The computation of distances were performed by using Dijkstra's algorithm, run from each vertex of the digraph, 
and terminating the search when all vertices within a given road network distance from the vertex were found. 
After each run of Dijkstra's algorithm, the out-neighbourhood of the corresponding vertex for the reachability digraph was recorded. 
As explained in Section~\ref{sec-domsets}, in the context of refuelling facilities and our definition of the $k$-dominating sets, the direction of all arcs in the original reachability digraph was reversed.

Dijkstra's algorithm using a binary heap based priority queue has a running time of $O((n+m)\log n)$ (e.g., see Bast et al. \cite{BH15}). This algorithm was called for each vertex of the road network digraph as the source vertex, giving a total running time of $O(n(n + m)\log n)$ for computing the corresponding reachability digraph. 
Notice that, as mentioned in Section~\ref{sec:reachability}, if energy consumption is used instead of road segment lengths for the arc weights in the road network digraph, negative arc weights are possible. 
This means that, in general, Dijkstra's algorithm cannot be used to compute reachability digraphs (e.g., see \cite{BG2009}).
In our experiments, the CPU time required to compute the road network digraphs varied from about one minute for small-size road networks to tens of minutes for large scale road networks.
For the obtained road network digraphs, the CPU time required to compute the corresponding reachability digraphs varied from seconds for small-size road networks to 2--3 days
for large scale road networks.

Our heuristic algorithms were tested on a number of reachability digraphs derived from road networks of the city of Birmingham and West Midlands conurbation (England).
These road networks were retrieved from OpenStreetMap database project \cite{OSM}. 
Each road network consists of all roads contained within a square bounding box centred at Birmingham New Street train station, the exact coordinates of the box centre are 
at $(52.478691, -1.899849)$.
The left diagram of Figure \ref{fig:Bmap.9} shows one such bounding box with a side length of $825\,\mathrm{m}$ meters, where the red points indicate the in-neighbourhood of the reachability digraph vertex (yellow point) using the reachability radius of 275 metres.

BG, DCG, and TCG were used alongside the randomized heurisitc based on the minimum, average, median, and maximum in-degree of the reachability digraph.
Since the randomized algorithm requires the in-degree parameter to be at least $k$, but the minimum in-degree of the reachability digraphs is usually equal to $0$,
the corresponding in-degree parameter value was set equal to $k$ instead (e.g., see Table~\ref{table:parameter}). 
If several in-degree parameter values were lower than $k$, their value were supposed to be set equal to $k$ too, which would lead to identical experiments.

The smaller scale experiments for this digraph type used digraphs derived from road networks with box side lengths of 0.75, 1, 1.25, 1.5, 1.75, and 2\,km,
using reachability radii of 200, 300, 325, 350, 375, and 400 metres, respectively.
Results for some of these experiments are presented in Table~\ref{RNEG1} (set size numbers in parentheses indicate a heuristic solution returned by Gurobi in 30 minutes; the best out-of-three results returned by the greedy heuristics are highlighted). 
For the heuristic results returned by the deterministic generic ILP solver (in parentheses in Table~\ref{RNEG1}), the optimal solution was not found even in 24 hours. 

Similarly to the small Erd\H{o}s--R\'enyi digraphs, Table~\ref{RNEG1} shows that the proposed greedy heuristics return a solution for the small size problem instances in a fraction of a second, while the solution quality is comparable to the exact or heuristic ILP solutions, after running the generic ILP solver on these small size instances for a much longer time. 
The best out-of-three greedy solution is within 0.5--10.8\% of the optimum for the $17$ cases where the optimum solution was found. 
Also, Table~\ref{RNEG1} shows that, for the small reachability digraphs, when $k>1$, TCG usually provides better results than the other two greedy heuristics, and the advantages of DCG and TCG over BG become more visible for the larger values of $k$. 

Out of the total 24 small size problem instances, the randomized heuristic solutions are better than the greedy heuristic solutions only for two road network instances (the 0.75\,km\,x\,0.75\,km road network, when $k=1$, and the 2\,km\,x\,2\,km road network, when $k=8$), and match the best out-of-three greedy solutions only in two other cases (the 0.75\,km\,x\,0.75\,km road network, when $k=4$, and the 1\,km\,x\,1\,km road network, when $k=1$). Therefore, we omitted these randomized heuristic results from the tables. 

%%%%%%%%%%
{
\begin{table*}
\footnotesize{
	\centering
	\begin{tabular}{|c|c|c|c|c|lr|lr|lr|lr|}
	\hline 
{\scriptsize Square} & {\scriptsize Radius,} & {\tiny \#Vertices,} & {\scriptsize \#Arcs,} & \multirow{2}{*}{$k$} & \multicolumn{2}{c|}{ILP}          & \multicolumn{2}{c|}{BG}       & \multicolumn{2}{c|}{DCG}     & \multicolumn{2}{c|}{TCG}     \\ \cline{6-13} 
{\scriptsize size} & {$r$} & {$n$} & {$m$} &					   & \multicolumn{1}{c|}{\scriptsize Size} & {\tiny Time\,(s)} & \multicolumn{1}{c|}{\scriptsize Size} & {\tiny Time\,(s)}  & \multicolumn{1}{c|}{\scriptsize Size} & {\tiny Time\,(s)}  & \multicolumn{1}{c|}{\scriptsize Size} & {\tiny Time\,(s)}  \\ 
\hline 
 \multirow{4}{*}{$(1\,\mathrm{km})^2$} &  \multirow{4}{*}{$300$\,m} & \multirow{4}{*}{614} & \multirow{4}{*}{8,546} &	$1$                  & \multicolumn{1}{r|}{61}   & 0.11   & \multicolumn{1}{r|}{\cellcolor{orange!25}66}   & 0.0057  & \multicolumn{1}{r|}{\cellcolor{orange!25}66}   & 0.0069 & \multicolumn{1}{r|}{67}   & 0.0063 \\ \cline{5-13} 
&&&&	$2$                  & \multicolumn{1}{r|}{108}  & 0.24    & \multicolumn{1}{r|}{118}  & 0.0085 & \multicolumn{1}{r|}{114}  & 0.011   & \multicolumn{1}{r|}{\cellcolor{orange!25}113}  & 0.0086 \\  \cline{5-13}
&&&&	$4$                  & \multicolumn{1}{r|}{195}  & 5.9     & \multicolumn{1}{r|}{220}  & 0.012  & \multicolumn{1}{r|}{211}  & 0.014  & \multicolumn{1}{r|}{\cellcolor{orange!25}210}  & 0.012  \\  \cline{5-13} 
&&&&	$8$                  & \multicolumn{1}{r|}{341}  & 170.8   & \multicolumn{1}{r|}{387}  & 0.017   & \multicolumn{1}{r|}{\cellcolor{orange!25} 355}  & 0.019  & \multicolumn{1}{r|}{358}  & 0.015  \\ 
\hline 
 \multirow{4}{*}{$(1.5\,\mathrm{km})^2$} &  \multirow{4}{*}{$350$\,m} & \multirow{4}{*}{1,387} & \multirow{4}{*}{25,958} &	
	$1$                  & \multicolumn{1}{r|}{108}  & 0.078   & \multicolumn{1}{r|}{\cellcolor{orange!25} 116}  & 0.021 & \multicolumn{1}{r|}{120}  & 0.036 & \multicolumn{1}{r|}{118}  & 0.043 \\ \cline{5-13}
&&&&	$2$                  & \multicolumn{1}{r|}{199}  & 4.63   & \multicolumn{1}{r|}{220}  & 0.032 & \multicolumn{1}{r|}{223}  & 0.042 & \multicolumn{1}{r|}{\cellcolor{orange!25} 218}  & 0.036 \\ \cline{5-13}
&&&&	$4$                  & \multicolumn{1}{r|}{(367)}  & 1,800     & \multicolumn{1}{r|}{412}  & 0.1  & \multicolumn{1}{r|}{394}  & 0.08 & \multicolumn{1}{r|}{\cellcolor{orange!25} 392}  & 0.052  \\ \cline{5-13}
&&&&	$8$                  & \multicolumn{1}{r|}{(638)}  & 1,800    & \multicolumn{1}{r|}{738}  & 0.075 & \multicolumn{1}{r|}{\cellcolor{orange!25} 684}  & 0.092 & \multicolumn{1}{r|}{686}  & 0.066  \\ 
\hline
 \multirow{4}{*}{$(2\,\mathrm{km})^2$} &  \multirow{4}{*}{$400$\,m} & \multirow{4}{*}{2,354} & \multirow{4}{*}{56,306} &	$1$                  & \multicolumn{1}{r|}{136}  & 0.099   & \multicolumn{1}{r|}{154}  & 0.044 & \multicolumn{1}{r|}{152}  & 0.055 & \multicolumn{1}{r|}{\cellcolor{orange!25} 148}  & 0.045 \\ \cline{5-13} 
&&&&	$2$                  & \multicolumn{1}{r|}{259}  & 7.4  & \multicolumn{1}{r|}{294}  & 0.064 & \multicolumn{1}{r|}{287}  & 0.076 & \multicolumn{1}{r|}{\cellcolor{orange!25} 283}  & 0.06 \\ \cline{5-13} 
&&&&	$4$                  & \multicolumn{1}{r|}{(485)}  & 1,800   & \multicolumn{1}{r|}{561}  & 0.11  & \multicolumn{1}{r|}{538}  & 0.14  & \multicolumn{1}{r|}{\cellcolor{orange!25} 525}  & 0.11  \\ \cline{5-13} 
&&&&	$8$                  & \multicolumn{1}{r|}{(899)}  & 1,800     & \multicolumn{1}{r|}{1,044} & 0.17  & \multicolumn{1}{r|}{\cellcolor{orange!25} 963}  & 0.19  & \multicolumn{1}{r|}{\cellcolor{orange!25} 963}  & 0.15  \\ \hline 
	\end{tabular}
	\caption{ILP solver and greedy heuristics results for $k$-dominating sets in small reachability digraphs.}
	\label{RNEG1}
	}
\end{table*}
}

The road network square bounding boxes used in the main experiments had side lengths of 10, 20, 30, 40, and 50\,km.
The reachability radii for these road networks were set equal to 3, 4, 5, 6, and 7\,km, respectively.
Table \ref{BMRN} provides further information about the corresponding road network and reachability digraphs. 
An attempt was made to run Gurobi on the reachability digraph corresponding to a road network square of side length 10 kilometres and the reachability radius of 3 kilometres, 
but the computer ran out of memory and was not able to produce any results.
An attempt was made to run computational experiments with our heuristics on the reachability digraph corresponding to a road network square of side length 60 kilometres and the reachability radius of 8 kilometres ($n=$\,300,252 vertices, 657,973 arcs in the road network digraph, $m=$\,1,808,497,368 arcs in the reachability digraph, $\delta^-=$\,0, $d^-_{\mathrm{ave}}=$\,6,023.27, $d^-_{\mathrm{med}}=$\,5,034,  $\Delta^-=$\,17,055). 
However, the computer did not have sufficient memory. 
Results of computational experiments with our heuristics for the other large size road networks are presented in Tables~\ref{RNG50} and \ref{table:random}. 
Table~\ref{table:parameter} gives values of the vertex in-degree parameter used by the randomized heuristic. 

%%%%%%%%%%%%%%%%%%%%%

\begin{table}[H]
	\centering
	\begin{tabular}{|c|c|r|r|r|}
	\hline
	\thead{\footnotesize Square road\\ \footnotesize network size}	& \thead{\footnotesize Reachabilty\\ \footnotesize radius, $r$} & \thead{\footnotesize \#Vertices,\\ $n$} 	& \thead{\footnotesize \#Arcs\\ {\scriptsize (road network digraph)}} & \thead{\footnotesize \#Arcs, $m$\\ {\scriptsize (reachability digraph)}}   \\ \hhline{|-|-|-|-|-|}
	$10\,\mathrm{km}\times10\,\mathrm{km}$	&  $3\,\mathrm{km}$              & 38,506   			& 85,450  		& 52,571,274    		\\ \hline
	$20\,\mathrm{km}\times20\,\mathrm{km}$	&  $4\,\mathrm{km}$              & 87,977   			& 193,919 		& 207,089,097   		\\ \hline
	$30\,\mathrm{km}\times30\,\mathrm{km}$	&  $5\,\mathrm{km}$              & 131,969  			& 289,952 		& 423,758,564   		\\ \hline
	$40\,\mathrm{km}\times40\,\mathrm{km}$	&  $6\,\mathrm{km}$              & 173,487  			& 380,125 		& 698,273,233   		\\ \hline
	$50\,\mathrm{km}\times50\,\mathrm{km}$	&  $7\,\mathrm{km}$              & 225,289  			& 492,880 		& 1,063,778,792 		\\ \hline
	\end{tabular}
	\caption{Birmingham and the West Midlands conurbation large road network and reachability digraph parameters.}
	\label{BMRN}
\end{table}

The best out-of-three results returned by the greedy heuristics are highlighted in Table~\ref{RNG50}. For comparison between the greedy heuristics, 
it can be seen that BG only matches the best solutions in two out of these twenty problem instances, and both for $k=1$, which can be explained by similarity of BG to DCG and TCG in this particular case. 
BG is still competitive for $k=1$, but for larger values of $k>1$, the advantages of DCG and TCG are more visible. 
Therefore, BG results could be ignored for these reachability digraphs.
DCG returns best solutions in 25\% of all the test instances, while TCG clearly performs better by finding the best out-of-three solutions in 80\% of the cases.
Notice that, for $k=1$, DCG would normally produce the same results as BG (TCG has the secondary selection criterion, which comes into play even when $k=1$). 
However, the random choice of a vertex among the equally most suitable candidates in iteration of DCG produces results slightly different from BG and 
introduces the option of running the algorithm several times to potentially obtain better results. 
The runtimes of greedy heuristics are always comparable on the same digraph instance and are reasonably short (less than $10$\,min for the reachability digraph on 225,289 vertices).

%%%%%%%%%
{
\begin{table*}
\normalsize{
	\centering
	\begin{tabular}{|c|c|lr|lr|lr|}
	\hline
{\footnotesize Square} & \multirow{2}{*}{$k$} & \multicolumn{2}{c|}{BG}      & \multicolumn{2}{c|}{DCG}    & \multicolumn{2}{c|}{TCG}    \\ \cline{3-8} 
{\footnotesize size} & & \multicolumn{1}{c|}{\footnotesize Size} & {\footnotesize Time\,(s)} & \multicolumn{1}{c|}{\footnotesize Size} & {\footnotesize Time\,(s)} & \multicolumn{1}{c|}{\footnotesize Size} & {\footnotesize Time\,(s)} \\  
\hline 
 \multirow{4}{*}{$(10\,\mathrm{km})^2$} & 	$1$           & \multicolumn{1}{r|}{88}   & 3.95   & \multicolumn{1}{r|}{91}   & 3.9   & \multicolumn{1}{r|}{\cellcolor{orange!25} 86}   & 4.36   \\ \cline{2-8}
&	$2$                  & \multicolumn{1}{r|}{151}  & 3.88   & \multicolumn{1}{r|}{149}  & 4.05    & \multicolumn{1}{r|}{\cellcolor{orange!25} 145}  & 4.61   \\ \cline{2-8}
&	$4$                  & \multicolumn{1}{r|}{260}  & 4.27   & \multicolumn{1}{r|}{259}  & 4.4   & \multicolumn{1}{r|}{\cellcolor{orange!25} 257}  & 4.92   \\ \cline{2-8}
&	$8$                  & \multicolumn{1}{r|}{452}  & 4.91   & \multicolumn{1}{r|}{450}  & 5.2   & \multicolumn{1}{r|}{\cellcolor{orange!25} 439}  & 5.79   \\ 
%%%%%%%
\hline
 \multirow{4}{*}{$(20\,\mathrm{km})^2$} & 	
 %%%
 	$1$                  	& \multicolumn{1}{r|}{179}  & 15.05   & \multicolumn{1}{r|}{\cellcolor{orange!25} 177}  & 15.48   & \multicolumn{1}{r|}{178}  & 18.14   \\ \cline{2-8}
&	$2$                  & \multicolumn{1}{r|}{305}  & 15.93   & \multicolumn{1}{r|}{\cellcolor{orange!25} 301}  & 16.61   & \multicolumn{1}{r|}{302}  & 19.43   \\ \cline{2-8}
&	$4$                  & \multicolumn{1}{r|}{542}  & 19.4    & \multicolumn{1}{r|}{515}  & 20.87   & \multicolumn{1}{r|}{\cellcolor{orange!25} 511}  & 23.12   \\ \cline{2-8}
&	$8$                  & \multicolumn{1}{r|}{961}  & 24.75   & \multicolumn{1}{r|}{902}  & 27.44   & \multicolumn{1}{r|}{\cellcolor{orange!25} 900}  & 28.97   \\ 
%%%%%
\hline
 \multirow{4}{*}{$(30\,\mathrm{km})^2$} & 	$1$                  & \multicolumn{1}{r|}{\cellcolor{orange!25} 232}  & 49.82   & \multicolumn{1}{r|}{235}  & 58.65   & \multicolumn{1}{r|}{\cellcolor{orange!25} 232}  & 50.81   \\ \cline{2-8}
&	$2$                  & \multicolumn{1}{r|}{398}  & 36.54   & \multicolumn{1}{r|}{400}  & 31.91    & \multicolumn{1}{r|}{\cellcolor{orange!25} 393}  & 36.55   \\ \cline{2-8}
&	$4$                  & \multicolumn{1}{r|}{708}  & 44.61   & \multicolumn{1}{r|}{\cellcolor{orange!25} 691}  & 39.77   & \multicolumn{1}{r|}{696}  & 49.5   \\ \cline{2-8}
&	$8$                  & \multicolumn{1}{r|}{1,275} & 49.55   & \multicolumn{1}{r|}{1,217} & 47.55   & \multicolumn{1}{r|}{\cellcolor{orange!25} 1,213} & 49.82   \\ 
%%%%%%
\hline
 \multirow{4}{*}{$(40\,\mathrm{km})^2$} &
 	$1$                  & \multicolumn{1}{r|}{\cellcolor{orange!25} 268}  & 98.52   & \multicolumn{1}{r|}{278}  & 120.9   & \multicolumn{1}{r|}{\cellcolor{orange!25} 268}  & 131.1   \\ \cline{2-8}
&	$2$                  & \multicolumn{1}{r|}{470}  & 145.1   & \multicolumn{1}{r|}{\cellcolor{orange!25} 468}  & 142.3   & \multicolumn{1}{r|}{\cellcolor{orange!25} 468}  & 143.9   \\ \cline{2-8}
&	$4$                  & \multicolumn{1}{r|}{853}  & 136.5   & \multicolumn{1}{r|}{828}  & 155.1   & \multicolumn{1}{r|}{\cellcolor{orange!25} 812}  & 151.9   \\ \cline{2-8}
&	$8$                  & \multicolumn{1}{r|}{1,540} & 177.2   & \multicolumn{1}{r|}{1,449} & 185.8   & \multicolumn{1}{r|}{\cellcolor{orange!25} 1,446} & 204.1   \\ 
%%%%%%
\hline 
 \multirow{4}{*}{$(50\,\mathrm{km})^2$} & 	$1$                  & \multicolumn{1}{r|}{338}  & 411.1   & \multicolumn{1}{r|}{334}  & 413.7   & \multicolumn{1}{r|}{\cellcolor{orange!25} 331}  & 371.2   \\ \cline{2-8}
&	$2$                  & \multicolumn{1}{r|}{570}  & 334.5   & \multicolumn{1}{r|}{571}  & 417.6   & \multicolumn{1}{r|}{\cellcolor{orange!25} 560}  & 398.9   \\ \cline{2-8}
&	$4$                  & \multicolumn{1}{r|}{982}  & 455.5   & \multicolumn{1}{r|}{\cellcolor{orange!25} 962}  & 401.2   & \multicolumn{1}{r|}{964}  & 449     \\ \cline{2-8} 
&	$8$                  & \multicolumn{1}{r|}{1,752} & 494.5   & \multicolumn{1}{r|}{1,695} & 543.1   & \multicolumn{1}{r|}{\cellcolor{orange!25} 1,689} & 494.1   \\ \hline
	\end{tabular}
	\caption{Greedy heuristics results for $k$-dominating sets in large reachability digraphs.}
	\label{RNG50}
	}
\end{table*}
}

%%%%%%%%%%%%%%%%
{
\begin{table*}
{
	\centering
	\begin{tabular}{|c|r|r|r|r|}
	\hline 
{\footnotesize Square} &  \multirow{2}{*}{$\delta^-$}   & \multirow{2}{*}{$d^-_{\mathrm{ave}}$}  & \multirow{2}{*}{$d^-_{\mathrm{med}}$}  & \multirow{2}{*}{$\Delta^-$}   \\ 
{\footnotesize size} &		&   &    &  \\ \hline 
%%%%%%%
{\footnotesize $(10\,\mathrm{km})^2$} & {$0\,(k)$}     & {1,365}    & {1,386}  & {2,566} \\ 
\hline 
{\footnotesize $(20\,\mathrm{km})^2$} &  {$0\,(k)$}     & {2,354}    & {2,514}  & {4,831} \\ 
\hline 
{\footnotesize $(30\,\mathrm{km})^2$} &  {$0\,(k)$}     & {3,211}    & {3,175}  & {6,957} \\ 
\hline 
{\footnotesize $(40\,\mathrm{km})^2$} &  {$0\,(k)$}     & {4,025}    & {3,707}  & {9,967} \\ 
\hline 
{\footnotesize $(50\,\mathrm{km})^2$} &  {$0\,(k)$}     & {4,722}    & {3,942}  & {13,362} \\ 
\hline 
	\end{tabular}
	\caption{Vertex in-degree parameter values of the large reachability digraphs used by the randomized heuristic.}
	\label{table:parameter}
	}
\end{table*}
}

%%%%%%%

Table~\ref{table:random} presents the computational results returned by the randomized heuristic, where the best out-of-four solutions are highlighted.
It can be seen that this heuristic was able to improve the greedy heuristics results (Table~\ref{RNG50}) for almost all problem instances,
except two problem instances, where it matched the best greedy solution (the 10\,km\,x\,10\,km road network, $k=2$)
and performed worse (the 20\,km\,x\,20\,km road network, $k=8$).
Although the flexibility of the randomized heuristic allowed us to improve the computational results in almost all of the cases, it can be seen that
the CPU time usage is about one order of magnitude higher than the corresponding CPU times of the greedy heuristics. This can be explained by the fact that the randomized heuristic 
was run 10 times on each problem instance, to select the best outcome (out of 10), while each greedy heuristic was run only once.

%%%%%%%%%%
{
\begin{table*}
\normalsize{
	\centering
	\begin{tabular}{|c|c|lr|lr|lr|lr|}
	\hline 
{\footnotesize Square} &  \multirow{2}{*}{$k$} & \multicolumn{2}{c|}{$p=p(\delta^-)$}       & \multicolumn{2}{c|}{$p=p(d^-_{\mathrm{ave}})$}       & \multicolumn{2}{c|}{$p=p(d^-_{\mathrm{med}})$}     & \multicolumn{2}{c|}{$p=p(\Delta^-)$}     \\ \cline{3-10} 
{\footnotesize size} & 				& \multicolumn{1}{c|}{\footnotesize Size} & {\footnotesize Time\,(s)} & \multicolumn{1}{c|}{\footnotesize Size} & {\footnotesize Time\,(s)}  & \multicolumn{1}{c|}{\footnotesize Size} & {\footnotesize Time\,(s)}  & \multicolumn{1}{c|}{\footnotesize Size} & {\footnotesize Time\,(s)}  \\ 
\hline 
\multirow{4}{*}{$(10\,\mathrm{km})^2$} &	$1$  					   
	                 & \multicolumn{1}{r|}{\cellcolor{orange!25}85}   & 61.29   & \multicolumn{1}{r|}{86}   & 54.93   & \multicolumn{1}{r|}{\cellcolor{orange!25}85}   & 53.8    & \multicolumn{1}{r|}{87}   & 51.85   		\\ \cline{2-10} 
&	$2$   & \multicolumn{1}{r|}{\cellcolor{orange!25}145}  & 67.14   & \multicolumn{1}{r|}{147}  & 58.92   & \multicolumn{1}{r|}{146}  & 57.36   & \multicolumn{1}{r|}{150}  & 56.74   		\\ \cline{2-10} 
&	$4$   & \multicolumn{1}{r|}{254}  & 66.75   & \multicolumn{1}{r|}{254}  & 62.55   & \multicolumn{1}{r|}{254}  & 62.98   & \multicolumn{1}{r|}{\cellcolor{orange!25}239}  & 73.86   		\\ \cline{2-10} 
&	$8$   & \multicolumn{1}{r|}{459}  & 70.57   & \multicolumn{1}{r|}{462}  & 75.93   & \multicolumn{1}{r|}{\cellcolor{orange!25}425}  & 89.83   & \multicolumn{1}{r|}{\cellcolor{orange!25}425}  & 84.16   		\\ 
\hline 
\multirow{4}{*}{$(20\,\mathrm{km})^2$} & 	
 	$1$          & \multicolumn{1}{r|}{174}  & 358.8   & \multicolumn{1}{r|}{175}  & 313.6   & \multicolumn{1}{r|}{176}  & 291.4   & \multicolumn{1}{r|}{\cellcolor{orange!25}173}  & 283.9   		\\ \cline{2-10}
&	$2$                  & \multicolumn{1}{r|}{307}  & 684.5   & \multicolumn{1}{r|}{300}  & 491.4   & \multicolumn{1}{r|}{296}  & 491.7   & \multicolumn{1}{r|}{\cellcolor{orange!25}295}  & 481.5   	\\ \cline{2-10}
&	$4$                  & \multicolumn{1}{r|}{545}  & 951.8   & \multicolumn{1}{r|}{\cellcolor{orange!25}492}  & 782.7   & \multicolumn{1}{r|}{\cellcolor{orange!25}492}  & 852.7   & \multicolumn{1}{r|}{519}  & 750.3   \\ \cline{2-10}
&	$8$                  & \multicolumn{1}{r|}{1,023} & 1,046    & \multicolumn{1}{r|}{921}  & 981.6   & \multicolumn{1}{r|}{924}  & 1,284    & \multicolumn{1}{r|}{\cellcolor{orange!25}904}  & 972.6   		\\ 
\hline
 \multirow{4}{*}{$(30\,\mathrm{km})^2$} & 	
	$1$                  & \multicolumn{1}{r|}{226}  & 1,489    & \multicolumn{1}{r|}{\cellcolor{orange!25}225}  & 1,114    & \multicolumn{1}{r|}{229}  & 1,249    & \multicolumn{1}{r|}{226}  & 1,068    	\\ \cline{2-10}
&	$2$                  & \multicolumn{1}{r|}{406}  & 1,739    & \multicolumn{1}{r|}{\cellcolor{orange!25}392}  & 1,126    & \multicolumn{1}{r|}{396}  & 1,331    & \multicolumn{1}{r|}{393}  & 1,031    	\\ \cline{2-10}
&	$4$                  & \multicolumn{1}{r|}{736}  & 1,819    & \multicolumn{1}{r|}{\cellcolor{orange!25}666}  & 1,353    & \multicolumn{1}{r|}{\cellcolor{orange!25}666}  & 1,347    & \multicolumn{1}{r|}{686}  & 1,019    	\\ \cline{2-10}
&	$8$                  & \multicolumn{1}{r|}{1,347} & 2,131    & \multicolumn{1}{r|}{\cellcolor{orange!25}1,181} & 1,371    & \multicolumn{1}{r|}{\cellcolor{orange!25}1,181} & 1,354    & \multicolumn{1}{r|}{\cellcolor{orange!25}1,181} & 1,199    	\\ 
\hline 
 \multirow{4}{*}{$(40\,\mathrm{km})^2$} & 	
 	$1$                  & \multicolumn{1}{r|}{286}  & 4,676    & \multicolumn{1}{r|}{269}  & 3,282    & \multicolumn{1}{r|}{267}  & 3,523    & \multicolumn{1}{r|}{\cellcolor{orange!25}263}  & 3,543    	\\ \cline{2-10}
&	$2$                  & \multicolumn{1}{r|}{499}  & 5,070    & \multicolumn{1}{r|}{470}  & 3,535    & \multicolumn{1}{r|}{469}  & 3,788    & \multicolumn{1}{r|}{\cellcolor{orange!25}463}  & 4,027    	\\ \cline{2-10} 
&	$4$                  & \multicolumn{1}{r|}{916}  & 5,499    & \multicolumn{1}{r|}{\cellcolor{orange!25}805}  & 3,659    & \multicolumn{1}{r|}{\cellcolor{orange!25}805}  & 3,945    & \multicolumn{1}{r|}{807}  & 3,991    	\\ \cline{2-10}
&	$8$                  & \multicolumn{1}{r|}{1,688} & 6,454    & \multicolumn{1}{r|}{\cellcolor{orange!25}1,436} & 3,804    & \multicolumn{1}{r|}{\cellcolor{orange!25}1,436} & 4,169    & \multicolumn{1}{r|}{\cellcolor{orange!25}1,436} & 4,269    		\\ 
\hline
 \multirow{4}{*}{$(50\,\mathrm{km})^2$} & 
	$1$                  & \multicolumn{1}{r|}{345}  & 7,701    & \multicolumn{1}{r|}{325}  & 5,552    & \multicolumn{1}{r|}{329}  & 6,238    & \multicolumn{1}{r|}{\cellcolor{orange!25}322}  & 6,637    	\\ \cline{2-10}
&	$2$                  & \multicolumn{1}{r|}{601}  & 9,778    & \multicolumn{1}{r|}{555}  & 5,817    & \multicolumn{1}{r|}{560}  & 5,624    & \multicolumn{1}{r|}{\cellcolor{orange!25}545}  & 5,868    	\\ \cline{2-10}
&	$4$                  & \multicolumn{1}{r|}{1,088} & 11,030   & \multicolumn{1}{r|}{966}  & 5,755    & \multicolumn{1}{r|}{971}  & 5,667    & \multicolumn{1}{r|}{\cellcolor{orange!25}914}  & 6,030    	\\ \cline{2-10}
&	$8$                  & \multicolumn{1}{r|}{1,999} & 11,430   & \multicolumn{1}{r|}{1,716} & 5,767    & \multicolumn{1}{r|}{1,713} & 5,884    & \multicolumn{1}{r|}{\cellcolor{orange!25}1,651} & 6,199    	\\ \hline
	\end{tabular}
	\caption{Randomized heuristic results for $k$-dominating sets in large reachability digraphs.}
	\label{table:random}
	}
\end{table*}
}

%%%%%%%%%
%%%%%%%%%%%%%%%%%%
\section{Conclusions}
\label{Conclusion}

This paper considers and emphasizes the use of digraphs for modelling problems in road networks, 
introduces the novel general 
concept of a reachability digraph corresponding to a road network,  
proposes and discusses modelling and optimization of facility locations in road networks by considering $k$-dominating sets in digraphs. 
Two new greedy heuristics for the optimization problem are developed and tuned by refining the selection criteria. 
They are computationally tested on two types of digraphs. 
We have shown and discussed the greedy heuristics performance with respect to some exact solutions (or heuristic solutions returned by a deterministic ILP solver)  
and each other.
Also, we use the probabilistic method to obtain an upper bound on the $k$-domination number of a digraph, tune the derived randomized heuristic, 
combining it with the most refined and effective greedy strategy to obtain better computational results in large-scale road networks. 
Frontiers of tractability for very large problem instances have been studied as well. 

This kind of modelling tools and solution methods should be relatively easy to integrate into outcome-driven decision support systems for regional planing and optimization (e.g., see \cite{NMP2024}).
Also, the heuristics can be used and implemented, e.g., in the NetworkX library of the programming language Python 
to improve and extend functionality of a primitive $1$-dominating set search algorithm to find smaller size multiple dominating sets in simple graphs as well as digraphs. 

The proposed heuristics are involved with some details, but have an advantage of being relatively simple, reasonably reproducible, and very efficient.
The randomized heuristic allows us to make the ``rigid" greedy strategies to become more flexible and effective.
For future research, it would be interesting to devise more subtle combinations of greedy and randomized techniques. 
We also plan to consider more subtle domination models in digraphs and more involved heuristic solution strategies, 
for example, applications and modifications of the local search. 
To help with exact solutions for small size problem instances in digraphs, one can consider devising customized deterministic algorithms (e.g., see \cite{Wendy1} for $1$-domination in simple graphs). 
It would be interesting to find a meaningful generalization or improvements for the bound of Theorem~\ref{ProbBound} and the corresponding randomized algorithm (Algorithm~\ref{RandAlg}) to make them more sensitive to some other parameters in digraphs, not just the minimum in-degree.

%%%%%%%
\section*{Acknowledgment}
\noindent Lukas Dijkstra acknowledges funding from the Maths DTP 2020, EPSRC grant EP/V520159/1.
%%%%%%%%%%

\end{document}